\PassOptionsToPackage{unicode=true,bookmarksopen=true,pdffitwindow=true,
  colorlinks=true,linkcolor=bluecite,citecolor=bluecite,
  urlcolor=bluecite,hyperfootnotes=false,
  pdfstartview={FitH},pdfpagemode=UseNone}{hyperref}
\PassOptionsToPackage{normalem}{ulem}
\documentclass[jqc,article,submit,moreauthors,pdftex]{Definitions/tsp}
\let\linenumbers\relax

\usepackage{xcolor} 
\usepackage{marginnote}
\input{Definitions/packageHack} 

\continuouspages{yes}
\firstpage{1} 
\pubvolume{1}
\issuenum{1}
\articlenumber{12345}
\pubyear{2025}
\copyrightyear{2025}
\datereceived{Day Month Year}
\dateaccepted{Day Month Year}
\dateonlinefirst{}
\datepublished{}

\Title{Investigating Techniques to Optimise the Layout of Turbines in a Windfarm using a Quantum Computer}

\Author{ James Hancock\textsuperscript{1}$^*$, Matthew Craven\textsuperscript{1}, Craig McNeile\textsuperscript{1}, Davide Vadacchino\textsuperscript{1}}

\AuthorNames{Matthew Craven, James Hancock, Craig McNeile, Davide Vadacchino}

\address{%
\textsuperscript{1}Centre for Mathematical Sciences, University of Plymouth, Plymouth, PL4 8AA, United Kingdom
}

\corres{Corresponding Author: James Hancock. Email: james.hancock@plymouth.ac.uk}

\secondnote{}
\abstract{    
This paper investigates Windfarm Layout Optimization (WFLO), where we formulate turbine placement considering wake effects as a Quadratic Unconstrained Binary Optimization (QUBO) problem. Wind energy plays a critical role in the transition toward sustainable power systems, but the optimal placement of turbines remains a challenging combinatorial problem due to complex wake interactions. With recent advances in quantum computing, there is growing interest in exploring whether hybrid quantum-classical methods can provide advantages for such computationally intensive tasks. We investigate
solving the resulting QUBO problem using the Variational Quantum Eigensolver (VQE) implemented on Qiskit's quantum computer simulator, employing a quantum noise-free, gate-based circuit model. Three classical optimizers are discussed, with a detailed analysis of the two most effective approaches: Constrained Optimization BY Linear Approximation (COBYLA) and Bayesian Optimization (BO). We compare these simulated quantum results with two established classical optimization methods: Simulated Annealing (SA) and the Gurobi solver. The study focuses on 4$\times$4 grid configurations (requiring 16 qubits), providing insights into near-term quantum algorithm applicability for renewable energy optimization.
}


\begin{document}

\section{Introduction}

The climate crisis and ongoing efforts to reduce emissions provide strong motivation for maximizing energy extraction from renewable sources. Utilizing these energy sources minimizes environmental impact and promotes a sustainable future.

One key renewable energy source is wind. There are many challenges to making it an economically viable and reliable energy source~\cite{veers2019grand}. In this paper, we focus on using quantum
computers to solve one specific problem towards maximizing the energy extracted from the wind. There are currently limitations on the available quantum computing hardware, which we will discuss later. Quantum computing applications in renewable energy have been reviewed in Ref. \cite{giani2021quantum,ajagekar2022quantum}. Recent work has specifically addressed quantum optimization methods~\cite{abbas2023quantum}.

Energy from the wind is extracted using wind turbines. Typically,
wind turbines are combined to form windfarms. The spatial layout of the turbines in the windfarm will change the maximum amount of
energy that the windfarm can produce. The use of optimization techniques for wind farm layout optimization dates back decades \cite{MOSETTI1994105}, with modern approaches continuing to develop \cite{manikowski2021multi}.

The exponential growth in optimization complexity with increasing turbine count motivates the development of improved algorithms for optimal turbine placement.  Recently, one form of the Windfarm Layout Optimization (WFLO) problem has been rewritten as a Quadratic Unconstrained Binary Optimization (QUBO) problem~\cite{senderovich2022exploiting} that can be solved on a quantum computer.

There are currently two dominant types of quantum computers. One is based on quantum circuits. For example, IBM and Google have constructed quantum computers based on this paradigm. One way to solve QUBO problems using a quantum circuit computer is to use a Variational Quantum Eigensolver (VQE).

Other types of quantum computers are Quantum Annealers (QAs)~\cite{QAoverview}.  QAs are built to solve QUBO problems.
D-WAVE's Advantage is the largest real-world QA, having 5000+ qubits
(although not all connected)~\cite{DWAd}. A new system currently under development, promising over 7000 qubits~\cite{DWAd2}, is expected to be available in the coming years. Fujitsu’s Digital Annealer can also solve QUBO problems using specialized hardware~\cite{matsubara2020digital}.

In this paper, we investigate using quantum optimization algorithms to solve the WFLO problem mapped to a QUBO problem using the \texttt{qiskit} package~\cite{Qiskit} from IBM, running on a classical simulator. We employ the Variational Quantum Eigensolver (VQE), an approach specifically designed for the Noisy Intermediate-Scale Quantum (NISQ) era~\cite{preskill2018quantum}. This work is a necessary first step to solving the WFLO problem using quantum computers.  

Apart from the importance of solving the WFLO problem to
produce more electricity from a windfarm, there are other motivations
for this study. QUBO problems are a subset of combinatorial optimization problems. Many optimisation problems can be mapped to QUBO. For example, the Maximum Cut (MAXCUT) problem from graph theory can be mapped to a QUBO. The QUBO problems from different applications may be easier or harder to solve using different methods, so it is important to find empirical evidence for the performance required for the solution in these different difficulty settings~\cite{abbas2024challenges}. Our contribution is to investigate solving the QUBO problem mapped from a WFLO using a simulator of a circuit-based quantum computer.

QUBO problems are one of the main problems that can
be solved by annealers such as those sold by D-Wave and Fujitsu~\cite{zahedinejad2017combinatorial}. The solution of WFLO problems may be an interesting test case for comparing the performance of adiabatic and circuit-based quantum computers. Previous work has shown the difficulty in comparing the performance of classical computers and adiabatic annealers in solving QUBO problems~\cite{albash2018demonstration,pearson2019analog,yaacoby2022comparison,martin2015unraveling}.

In this work, we have used the circuit model of quantum computing as the
basis of the calculation. Namely, \texttt{qiskit}'s gate-based quantum circuit model with superconducting qubit architectures, as typically employed in IBM quantum systems. A key challenge in current quantum computing is the absence of quantum error correction. This necessitates algorithms with inherent resilience to decoherence. In practice, NISQ era algorithms typically adopt hybrid quantum-classical architectures to mitigate these limitations. We limit ourselves to a regime where we do not consider the quantum noise effects, such as decoherence. Instead, we tackle the problem in a statistically noisy setting, providing a necessary baseline for future work incorporating quantum noise effects.

\begin{figure}
    \centering
    \includegraphics[width=1.0\textwidth]{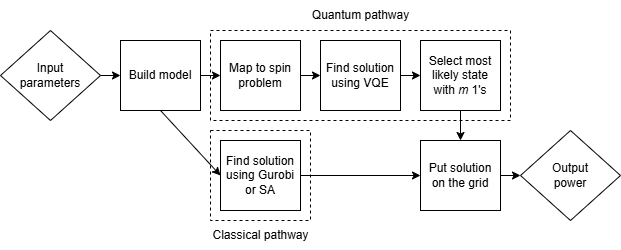}
    \caption{\corr{Flowchart of overall process of WFLO.}}
    \label{fig:flowchart}
\end{figure}

The workflow diagram in \cref{fig:flowchart} shows our overall process, where the quantum and classical pathways are highlighted.

\section{Windfarm Layout Optimization}

WFLO involves selecting turbine locations to maximize power output under specific constraints. These include: a maximum number of turbines, and enforcing a minimum separation distance between turbines due to rotor interference. We assume turbines can only be placed on discrete grid points.

Turbines positioned downstream (i.e., behind other turbines relative to wind direction) experience reduced power output compared to those exposed to undisturbed wind. This wind speed velocity deficit, known as the turbine wake, creates the fundamental challenge in WFLO: minimizing energy generation losses from wake interactions.

The Sum-of-Squares (SS) model for wake speeds~\cite{SSproceedings}
best captures the effect of these wakes. However, the formulation leads to intractable optimization problems. To combat this, we use a simplified model, the Linear Superposition of Wakes (LS) \cite{Donovan2005WindFO}. This allows us to map the problem to a QUBO problem, which can then, in turn, be optimized using various quantum (and classical) methods.

\section{Windfarm layout as a Quadratic Unconstrained Binary
  Optimization problem}

A QUBO problem is defined as
\begin{equation}
    \underset{x}{\text{argmin }}f_Q(x)\text{,}
\end{equation}
where
\begin{equation}\label{eq:xQx}
    f_Q(x) := x^TQx = \sum_{i,j=0}^{q}Q_{ij}x^ix^j
\end{equation}
for $x\in\mathbb{B}^q=\{0,1\}^q$.

Many combinatorial optimization problems can be formulated as QUBO
problems. QUBO is a class of Nondeterministic Polynomial Time (NP)-hard problems~\cite{QUBOsurvey}. While quantum computers offer ways to find high-quality solutions to QUBO problems, finding guaranteed optimal solutions is not expected to be in the Bounded-Error Quantum Polynomial Time (BQP) complexity class for general cases. QUBO problems, by definition, are formulated in an unconstrained way. However, constraints can be included by introducing penalties in the form of large additive constants when the conditions are violated. We follow the mapping of the WFLO problem to a QUBO problem that was developed by Senderovich et al in Ref. \cite{senderovich2022exploiting}.

We model the wind farm terrain as a square grid with side length $l_{\text{grid}}$, allowing for $l_{\text{grid}}^2$ potential turbine placement locations. These will be labelled as they are in \cref{fig:gridlabelling}. The indicator of the presence of a turbine on site $q$ is $x^q$, where $q \in \{1,...l_{grid}^2\}$. Cartesian coordinates of the grid are shown in \cref{fig:gridcord}.

\begin{figure}[H]
    \centering
    \begin{minipage}{0.45\textwidth}
        \centering
        \begin{tikzpicture}
        \draw[step=1cm,black,very thin] (0,0) grid (4,4);
        \draw (0.5,3.5) node{1};
        \draw (0.5,2.5) node{5};
        \draw (0.5,1.5) node{9};
        \draw (0.5,0.5) node{13};
        \draw (1.5,3.5) node{2};
        \draw (1.5,2.5) node{6};
        \draw (1.5,1.5) node{10};
        \draw (1.5,0.5) node{14};
        \draw (2.5,3.5) node{3};
        \draw (2.5,2.5) node{7};
        \draw (2.5,1.5) node{11};
        \draw (2.5,0.5) node{15};
        \draw (3.5,3.5) node{4};
        \draw (3.5,2.5) node{8};
        \draw (3.5,1.5) node{12};
        \draw (3.5,0.5) node{16};
        \end{tikzpicture}
        \caption{Labelling of sites on a $l_{grid}=4$ windfarm grid.}
        \label{fig:gridlabelling}
    \end{minipage}\hfill
    \begin{minipage}{0.45\textwidth}
        \centering
        \begin{tikzpicture}
        \draw[step=1cm,black,very thin] (0,0) grid (4,4);
        \draw (0.5,3.5) node{(1,1)};
        \draw (0.5,2.5) node{(2,1)};
        \draw (0.5,1.5) node{(3,1)};
        \draw (0.5,0.5) node{(4,1)};
        \draw (1.5,3.5) node{(1,2)};
        \draw (1.5,2.5) node{(2,2)};
        \draw (1.5,1.5) node{(3,2)};
        \draw (1.5,0.5) node{(4,2)};
        \draw (2.5,3.5) node{(1,3)};
        \draw (2.5,2.5) node{(2,3)};
        \draw (2.5,1.5) node{(3,3)};
        \draw (2.5,0.5) node{(4,3)};
        \draw (3.5,3.5) node{(1,4)};
        \draw (3.5,2.5) node{(2,4)};
        \draw (3.5,1.5) node{(3,4)};
        \draw (3.5,0.5) node{(4,4)};
        \end{tikzpicture}
        \caption{Coordinates of sites on a $l_{grid}=4$ windfarm grid.}
        \label{fig:gridcord}
    \end{minipage}
\end{figure}

A wake can be defined by three parameters: the angle $\alpha$ (degrees) that the wind is coming from with respect to the west counted clockwise, $x$, the maximum distance that is affected by the wake and $r$ is the radius with which it spreads out per unit distance, away from the turbine causing the wake. These physical factors are problem-specific. For our model, we say that if a grid location is at all within the wake, it is entirely within its wake. We denote a wake starting at position $i$ as $w(i,x,r;\alpha)$.

\begin{figure}[H]
    \centering
    \begin{minipage}{0.45\textwidth}
        \centering
        \includegraphics[width=\linewidth]{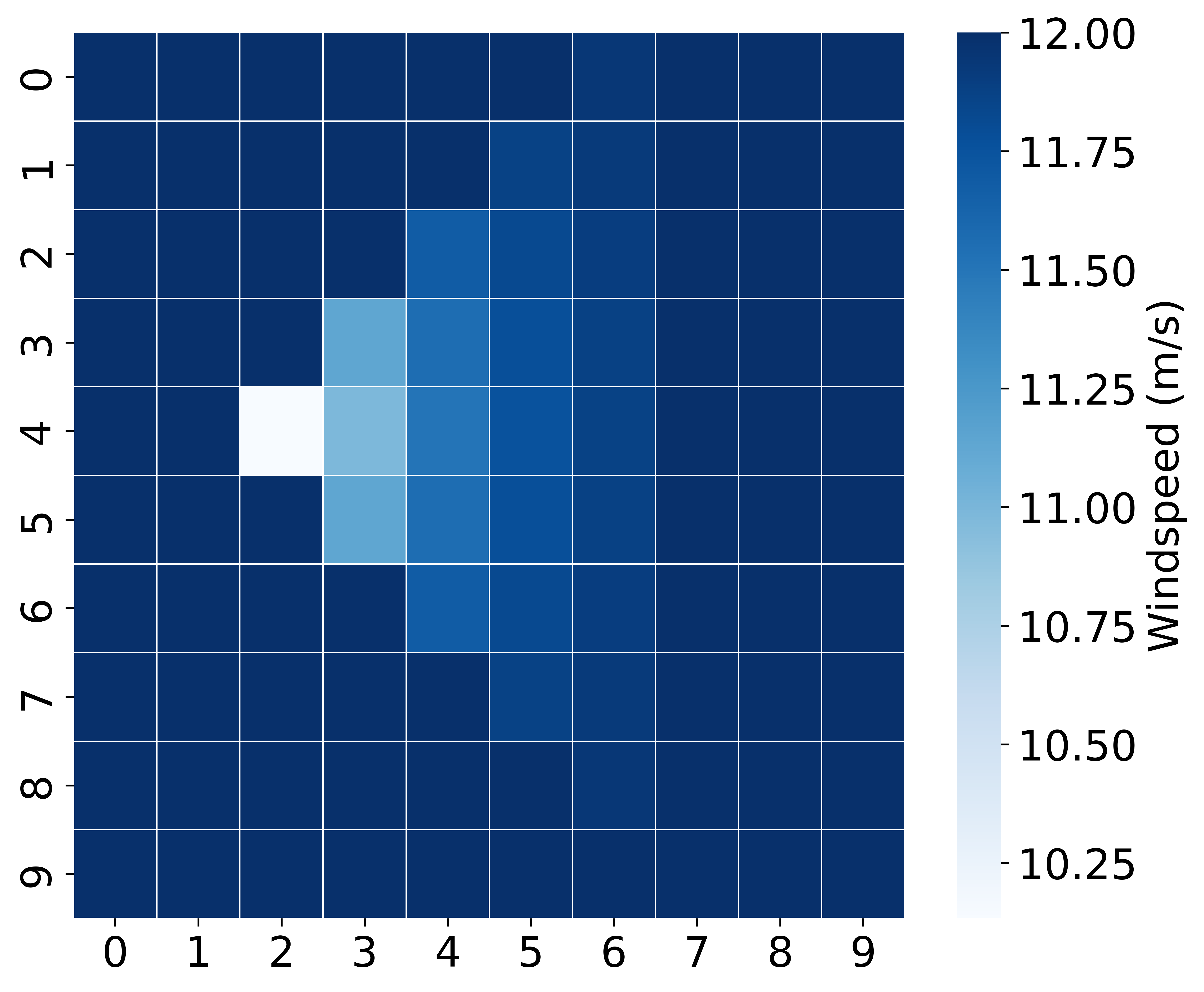}
        \caption{Turbine on location 15 or (2,4), $D=\{0,12ms^{-1},1\}$.}
        \label{fig:sub1A}
    \end{minipage}\hfill
    \begin{minipage}{0.45\textwidth}
        \centering
        \includegraphics[width=\linewidth]{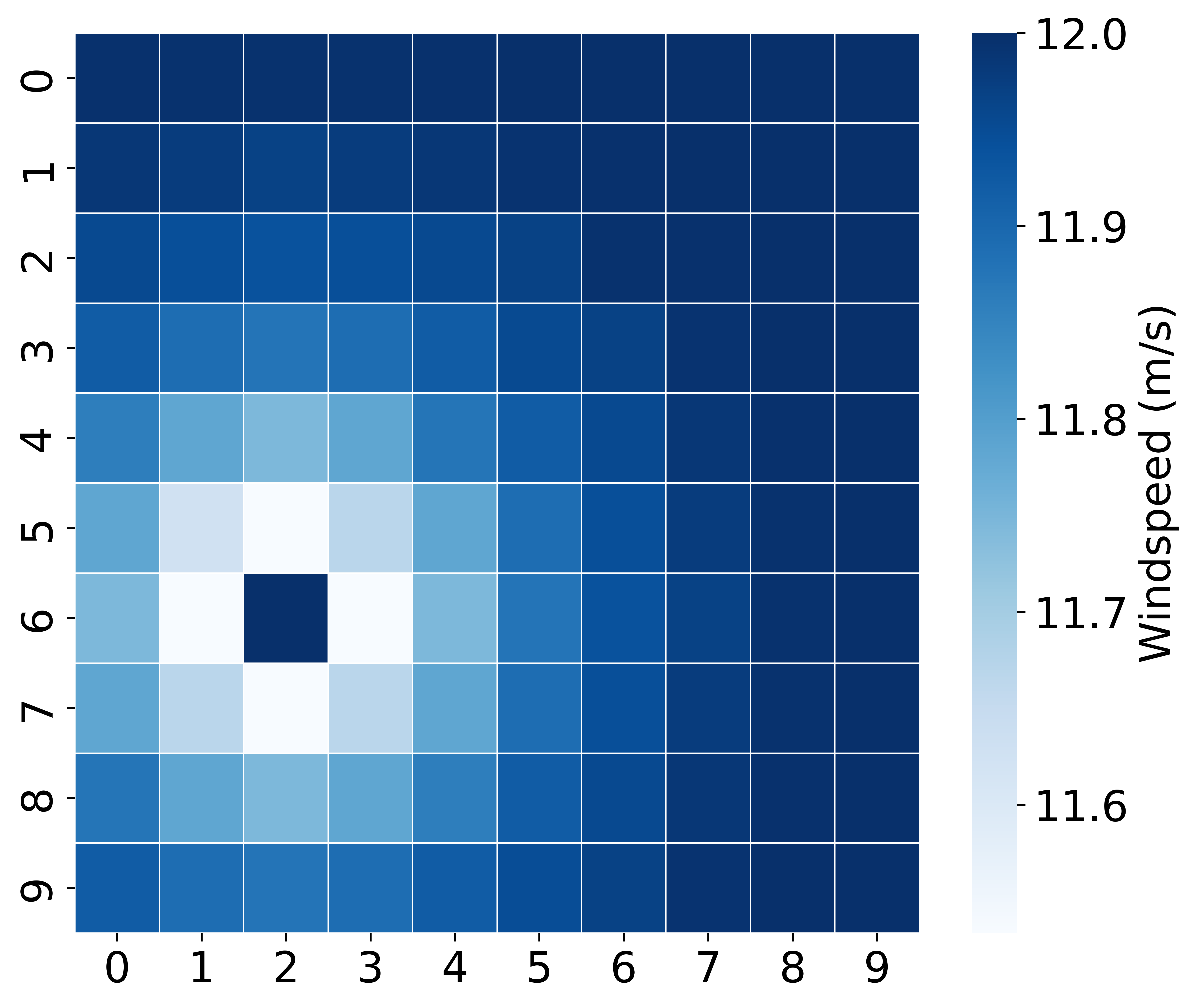}
        \caption{Turbine on location 27 or (2,6), $D=$ Second Mosetti.}
        \label{fig:sub2A}
    \end{minipage}
\end{figure}

\begin{figure}[H]
    \centering
    \begin{minipage}{0.45\textwidth}
        \centering
        \includegraphics[width=\linewidth]{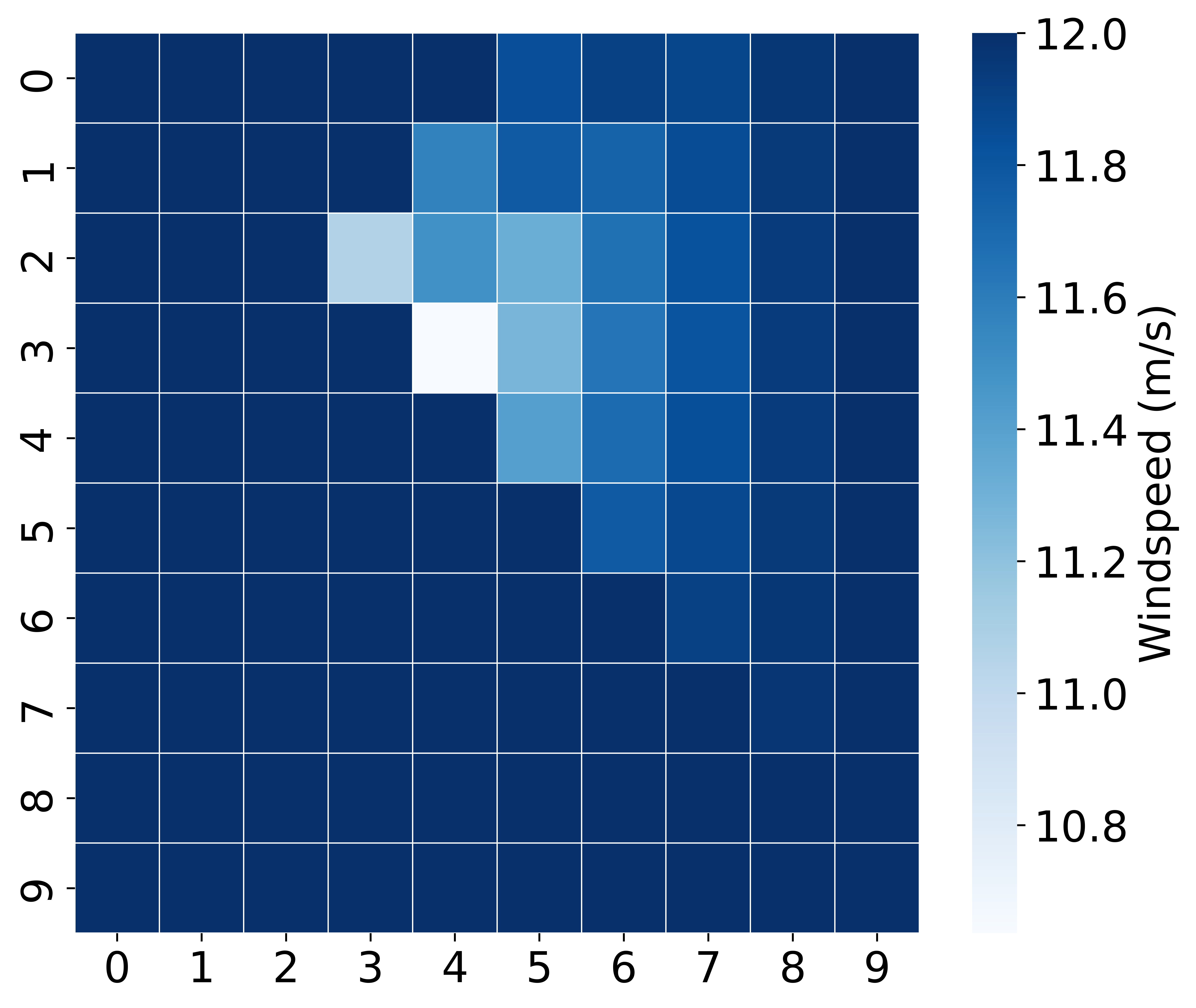}
        \caption{Turbine on location 23 or (2,2) and 34 or (3,3), $D=\{0,12ms^{-1},1\}$.}
        \label{fig:sub1B}
    \end{minipage}\hfill
    \begin{minipage}{0.45\textwidth}
        \centering
        \includegraphics[width=\linewidth]{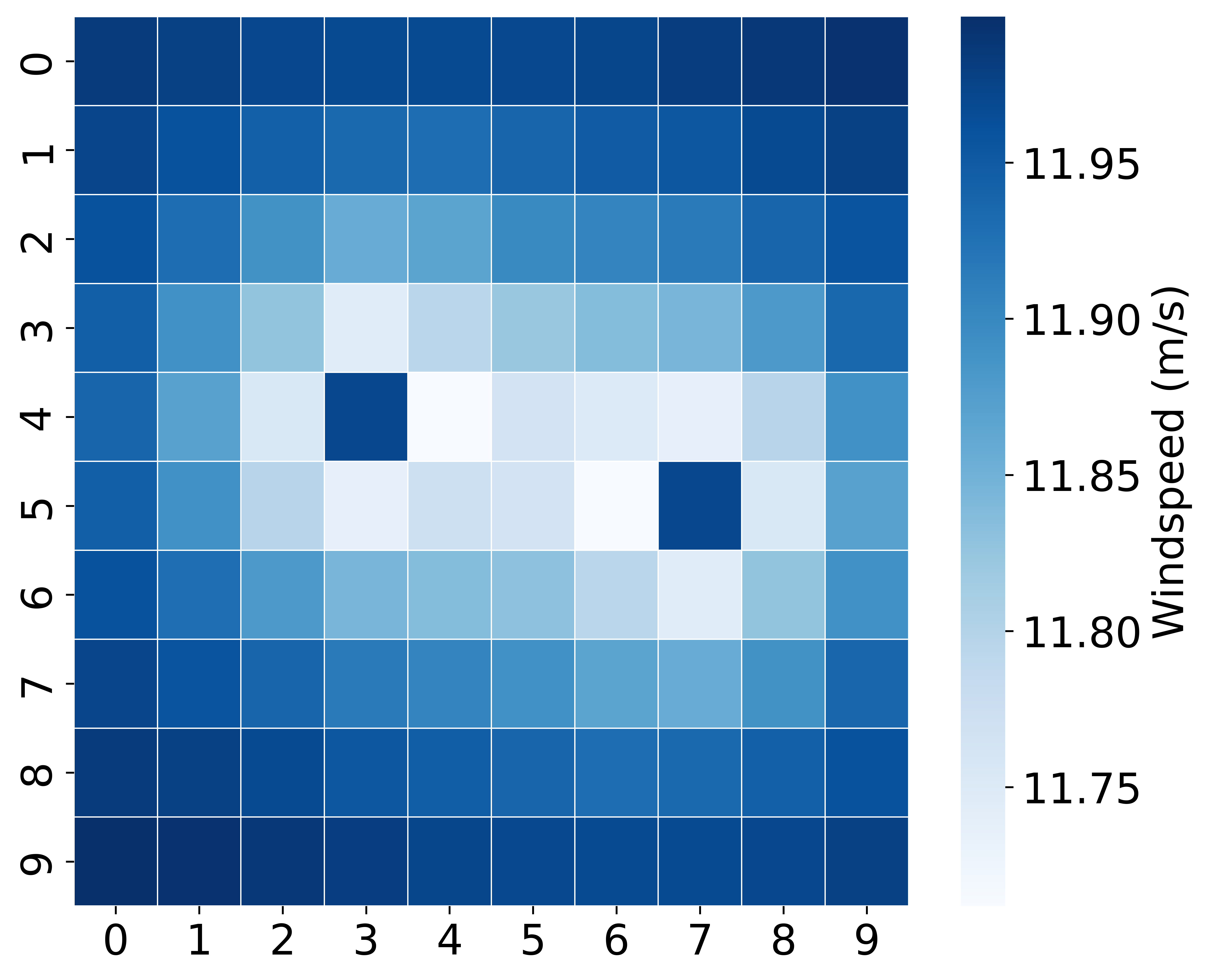}
        \caption{Turbine on location 35 or (3,4) and 76 or (7,5), $D=$ Second Mosetti.}
        \label{fig:sub2B}
    \end{minipage}
\end{figure}
To visualize the wakes from some turbines in the wind, they are plotted
on a $l_{grid} = 10$ grid (larger than that used in the simulations)
with the parameters: $x = 3$, $r=1$. \cref{fig:sub1A} and \cref{fig:sub2A} show a configuration of the wakes from one turbine. \cref{fig:sub1B} and \cref{fig:sub2B} show a configuration of the wakes from two turbines. If one of the turbines is in the wake of another turbine, then it will see a reduced wind velocity and thus produce less power.

\subsection{Model}

To ensure that the model is more realistic, the system includes several wind arrangements. Each possible wind arrangement is defined as $d:=\{\alpha_d,v_d,p_d\}$. $\alpha_d$ is the angle at which the wind comes in, $v_d$ is the \textit{free wind speed}, i.e., the wind speed that powers a turbine not in the wake of another. $p_d$ is the probability that we would expect to encounter this wind arrangement. We denote the set of possible wind arrangements in our system as $D$, which we call the \emph{wind regime}. To clarify this, $D$ is a collection of $d$'s, with $\sum_{d\in D}p_d = 1$.

The particular model that we are using is the LS of wakes. This model is useful as it can be transformed into a QUBO problem. The power output for a system of $q$ sites within this model is computed as:
\begin{equation}\label{eq:ELS}
    E^{\text{LS}} = \sum_{d\in D}\sum_{i=1}^q p_d\left[\frac{1}{3}v_d^3 - \sum_{j\in w_i}\frac{1}{3}\left(v_d^3 - u_{ij}^3\right)\right],
\end{equation}
where $w_i$ is shorthand for the wake caused by turbine $i$, and $u_{ij}$ is the reduced wind speed at site $j$, caused by being in the wake of $i$. If $j$ is in the wake of $i$, $u_{ij}$ can be calculated as
\begin{equation}\label{eq:red}
    u_{ij} = v_d\left(1-\frac{2a}{\left(1+\alpha_T(\delta/r)^2\right)^2}\right).
\end{equation}
This is the Jansen wake model \cite{khanali2018optimizing}, where $a$ is the axial induction factor, currently set to 0.1, and $\delta=||i-j||_2$ is the distance between the turbines. By controlling the wake's downstream length ($x$) and its radial expansion rate ($r$), we compute $\alpha_T$ using \cref{eq:alp}:
\begin{equation}\label{eq:alp}
    \alpha_T = \frac{1}{x}\left(r - r_t\right),
\end{equation}
where $r_t$ denotes the turbine radius. In this work, we set $r_t=0.33$ (in terms of grid boxes). We note that $r_t$ and $a$ from \cref{eq:red} carry units; their values are dependent on the physical features of the turbines. The selected values are arbitrary and do not correspond to any physical system, as this represents a simplified theoretical model.

The QUBO formulation for this problem is then:
\begin{equation}
    \underset{x}{\text{argmin }}(-f(x)),
\end{equation}
\begin{equation}\label{eq:Energy}
    f(x) = \sum_{d\in D}\sum_{i=1}^m p_d\left[\frac{1}{3}v_d^3x^ix^i - \sum_{j\in w_i}\frac{1}{3}\left(v_d^3 - u_{ij}^3\right)x^ix^j\right].
\end{equation}

This represents the unconstrained problem formulation, where the optimal solution trivially places a turbine at every available site. To enforce practical constraints, we introduce the energetic penalty:
\begin{equation}\label{eq:con}
    g(x;\lambda_1,\lambda_2) = \lambda_1\left(\sum_{i=1}^{l_{grid}^2}x^i - m\right)^2 + \lambda_2\sum_{||i-j||_2<\xi}x^ix^j
\end{equation}
The term weighted by $\lambda_1$ in \cref{eq:con} limits the number of total turbines to $m$; the second term pertains to the minimum distance between the turbines being $\xi$. The values of $\lambda_1,\lambda_2$ have to be large enough so that the constraints are met. The full QUBO problem is:
\begin{equation}\label{eq:Qfg}
    \underset{x}{\text{argmin }}(-f(x)+g(x;\lambda_1,\lambda_2))
\end{equation}

The problem from \cref{eq:Qfg} can then be written in the form of a weight matrix Q from \cref{eq:xQx}:
\begin{equation}\label{eq:Q}
    Q_{ij} = \begin{cases}
        -\frac{1}{3}\sum_d p_d v_d^3 + \lambda_1(1-2nm)& \text{if $i=j$}\\
        \\
        -\frac{1}{3}\sum_d p_d \left(v_d^3-u_{ij}^3\right) + 2\lambda_1 & \text{if $i\neq j$}
    \end{cases}
\end{equation}
Here, we have neglected the proximity constraints term as we do not include this in our simulations. When the problem is formulated, it is a fully dense $Q$ matrix. We note here that $q = l_{grid}^2$ is the number of variables/qubits.

\section{Classical methods}

As a baseline comparison to the use of quantum computers to solve QUBO problems, we also investigated two classical optimization techniques. 

Gurobi is a suite of optimization solvers compatible with multiple programming languages, capable of solving both linear and quadratic problems. The software automatically selects algorithms from its toolkit, including simplex and parallel barrier methods, based on problem characteristics. For further details, see Ref.~\cite{gurobis}. In this work, we used Gurobi to solve our QUBO problem, which is a special case of quadratic optimization. Gurobi has been used as a classical benchmark for solving QUBO problems in comparative studies with quantum and digital annealers \cite{cseker2022digital,nakano2022diverse}.

Simulated Annealing~\cite{kirkpatrick1983optimization} (SA) is a standard approach to finding the optimum of a function that has many local minima in a large search space. The algorithm is especially effective for discrete space exploration. It navigates the solution space through controlled random fluctuations, where a temperature-dependent probability function governs state transitions. SA has been used to find solutions to various NP-complete combinatorial problems such as the Traveling Salesman Problem \cite{SATSP}, Minimum Linear Arrangement \cite{SAMLA}, and instances of the Packing Problem \cite{SAPP}.

\section{Variational Quantum Eigensolver}

The VQE
is a hybrid quantum-classical algorithm for finding the lowest
eigenvalue of a Hamiltonian that is in the form of a Pauli string~\cite{VQEorig}. For further details about the VQE, see
Ref.~\cite{VQE}. The Variational Quantum Eigensolver (VQE) has become a pivotal algorithm for the NISQ era. The approach employs a parameterized quantum circuit to compute the Hamiltonian expectation value with respect to parameters $\theta$, coupled with a classical optimization routine that minimizes this value through iterative $\theta$ updates.

The VQE uses the inequality
\begin{equation}\label{eq:VQEE}
    E_{min}^H \leq \langle 0|U^\dagger(\theta)HU(\theta)|0\rangle,
\end{equation}
and the variational principle to find a tight upper bound on the lowest eigenvalue, $E_{min}^H$, of a Pauli string matrix, $H$, which is defined as
\begin{equation}\label{eq:Pstring}
    H = \sum_\gamma h_\gamma P_\gamma
\end{equation}
\begin{equation}
    P_\gamma = \bigotimes_{i=0}^{q}\sigma_{m_i},
\end{equation}
where $m_i\in\{0,3\}$ tells us the Pauli matrix and $q$ is the number of qubits. The two Pauli matrices that we make use of in this problem formulation are:
\begin{equation}
    \sigma_0 = \begin{pmatrix}
        1 & 0\\
        0 & 1
    \end{pmatrix} = I_2,
\end{equation}
and
\begin{equation}
    \sigma_3 = \begin{pmatrix}
        1 & 0\\
        0 & -1
    \end{pmatrix}.
\end{equation}
The parameter values are then chosen by a classical optimization routine to find the value for
\begin{equation}
    \underset{\theta}{\text{min }}\langle 0|U^\dagger(\theta)HU(\theta)|0\rangle
\end{equation}

There are two main features of the VQE that we must design and control for efficient implementation. These are: the design of the parameterized circuit (the \textit{ansatz}) and the choice of optimization routine. Our choice of ansatz must create states that have sufficient overlap with the optimal state in the solution space. When using the VQE to solve QUBO problems, this is fairly trivial, as one parameter on each qubit would be sufficient (we are only looking for basis states). However, selecting an appropriate optimization routine presents a non-trivial challenge that critically impacts algorithm performance. In this work, we examine the effectiveness of three different optimization routines: Powell's Optimization (PO), Constrained Optimization BY Linear Approximation (COBYLA), and Bayesian Optimization (BO).

\subsubsection{Powell optimization}

PO is a widely known standard optimization routine for finding the global minimum of a function without the use of derivatives. It works well for the VQE (in small parameter spaces) but requires the system to use a high shot \footnote{Number of calls of the quantum circuit to gain one expected value evaluation.} count in order not to be too greatly affected by the quantum measurement noise. While this method is robust with high iterations for a black-box problem and does not use derivative information, it is poorly suited to the quantum formulation due to the sheer number of shots and iterations required.

For this work, we make use of \texttt{scipy.optimize.minimize}'s option \texttt{`powell'}.

\subsubsection{COBYLA}

COBYLA is a trust region-based, surrogate-assisted method for finding
the global minimum of a function. Similar to PO, COBYLA requires
sufficient shots in order not to be too greatly affected by the noise.
However, it is more resilient than PO. COBYLA has been used for the VQE
several times in the literature. Liu et al. \cite{COBYLA1} used it as a test for their Layer-VQE (L-VQE) approach to generating ansatz, and Tim Schwägerl, et al. \cite{COBYLA2} used it to minimize a QUBO problem for
reconstructing particle track results from a Large Hadron Collider (LHC). COBYLA has shown good promise for this black-box optimization. The lack of need for derivative information and the use of surrogate models make COBYLA fairly well-suited for the noisy environment in which we are working. Its trust-region approach provides some inherent noise robustness. COBYLA is efficiently implemented in Python, which makes it a useful choice. 

For this work, we make use of \texttt{scipy.optimize.minimize}'s option \texttt{`COBYLA'}.

\subsubsection{Bayesian Optimization}

BO is a surrogate-assisted method for finding the global minimum of a
function, which models the underlying function as a Gaussian Process
Regression (GPR) and then updates this model based on samples,
according to Bayes' theorem \cite{McClean:2015vup,Klco:2018kyo}. BO has been extensively studied in the field of hyperparameter optimization for machine learning (e.g., Refs. \cite{PBh,BOa,AfML}). BO is well suited to the noisy environment of quantum optimization, as it takes this variance into account when building surrogate Gaussian Process (GP) models. BO has a high computational cost, especially as the number of iterations grows.

For this work, we use a BO that we have coded. The definitions that we use are the same as presented in Ref. \cite{KJ}, except for the choice of \textit{acquisition function}. We use the periodic kernel, for $K$ parameters, defined as,
\begin{equation}
    k^P(\theta^1,\theta^2) = \sigma^2 \prod_{i=1}^K\exp\left[\frac{-2}{l^2}\sin^2\left(\pi \left|\frac{\theta^1_i-\theta^2_i}{p}\right|^2\right)\right],
\end{equation}
where $\sigma$ is the expected variance of the GP, $l$ is the length scale over which we would expect data points to be correlated, and $p$ is the period of the underlying function. 

We make use of Expected Improvement (EI) as the acquisition function. If we define our best guess so far to be $E_{min} = \min(E_1,...E_n)$, EI is defined as
\begin{equation}
    a_{EI}(\theta) = \mathbb{E}_{\theta}[\max(0,E_{min}-E(\theta))]
\end{equation}
Where $\mathbb{E}_{\theta} [\cdot]$ is evaluated over all the possible $E(\theta)$ from the surrogate model.

Expected Improvement can be used in its explicit closed form, which is,
\begin{equation}
\begin{aligned}
    a_{EI}(\theta) = &\left(E_{min}-E(\theta)\right)\Phi\left(\frac{E_{min}-E(\theta)}{\Delta E(\theta)}\right)\\& + \Delta E(\theta) \phi\left(\frac{E_{min}-E(\theta)}{\Delta E(\theta)}\right).
\end{aligned}
\end{equation}
Where $\Phi$ is the Normal cumulative distribution function and $\phi$ is the corresponding probability density function. We find the minimum of $a_{EI}$ using COBYLA.

\subsection{Mapping QUBO to the VQE}\label{sec:Q2VQE}
We can use the VQE to solve QUBO problems using the transformation
\begin{equation}
    x^i \mapsto \frac{1}{2}\left(\sigma^i_0+\sigma_3^i\right),
\end{equation}
where $\sigma_3^i$ denotes the Pauli-Z spin matrix acting on the $i^{th}$ qubit. We then find the groundstate eigenpair of the Hamiltonian
\begin{equation}
    H = \sum_{i,j=1}^q Q_{ij}\frac{1}{4}\left((\sigma^i_0+\sigma_3^i)(\sigma^j_0+\sigma_3^j)\right).
\end{equation}
Once found, we take the parameters that produce this groundstate, measure the state in the $\sigma_3^{\otimes q}$ basis, and the state with the highest likelihood of being measured, which also meets our constraints, is our solution. 

Since our Hamiltonians only consist of $\sigma_0$ and $\sigma_3$ terms, they are diagonal; hence, we only need one parameter per qubit to map a sufficient amount of the space to receive the optimum. However, to aid the classical optimizer in the noisy space of quantum measurements, $q$ layers of rotation and entangling Controlled-$X$ or Controlled-$NOT$ ($CNOT$) gates will be used. 

We note here that due to the density of the weight matrix $Q$ defined in \cref{eq:Q}, the Hamiltonian is fully dense, containing $q^2$ nonzero terms. This means that this is a worst-case scenario in terms of stress on the VQE.

\subsection{Dimensionallity Expressivity Analysis}
In variational quantum circuits, Dimensionality Expressivity Analysis (DEA) \cite{Funcke_2021} serves as a key tool for assessing independence among quantum gate parameters. This method identifies whether gate parameters represent independent degrees of freedom or exhibit redundancy. Full details on DEA methodology are available in \cite{funcke2021dimensional}.

DEA is more applicable when looking at complex systems, i.e., when we are not only searching for basis states, but a specific subspace is required.

When performing this analysis on our circuit, we observe that all parameters remain non-redundant provided none are initialized at $(2k+1)\pi/4$ or $2k\pi/4$ for $k \in \{0,1\}$. To guarantee this condition, we initialize parameters by sampling uniformly from $[0, 2\pi)$ using \texttt{numpy.random.uniform}.

Further details about DEA are included in \cref{appen:DEA}.

\subsection{Conditional Value at Risk}
A significant advancement in applying VQE to combinatorial optimization involves adopting the Conditional Value at Risk (CVaR) measure from financial mathematics. While Ref.~\cite{CVaR} implemented CVaR-VQE using the sample mean, our work applies this approach to the quantum expectation value. See \cite{CVaR} for complete methodology and advantages of using CVaR.

For a Hamiltonian on which we carry out $K$ measurements (shots), our objective function terms are,
\begin{equation}
    \text{VQE} \rightarrow \sum_{\gamma=1}^K H_\gamma
\end{equation}
If we now order the measurements so that $H_{k}\leq H_{k+1}$, the CVaR adaptation is to introduce a new parameter: $0<\alpha\leq 1$ such that,
\begin{equation}
    \text{CVaR-VQE} \rightarrow \sum_{\gamma=1}^{\lceil \alpha K\rceil} H_\gamma
\end{equation}
This approach truncates measurements to include only the terms contributing most significantly to the minimization objective, specifically, the largest negative contributions. The CVaR method considers only the lowest $\alpha K$ measurements, where adjusting $\alpha$ enables faster convergence to the minimal energy state. However, very small $\alpha$ values may yield unstable results due to severe measurement reduction. The standard VQE is recovered when $\alpha=1.0$.

Note that the minimum value of CVaR-VQE is not necessarily equal to the true minimum of the VQE, but the correct solution state can be the same.

\section{Design of the tests of the algorithms}

We evaluated power production across methods using $l_{\text{grid}}=4$ grids, limited by computational constraints. Through exhaustive search, we determined the maximum achievable power for this grid size. Given the WFLO problem's numerous local optima, each algorithm was executed 36 times with distinct initial conditions. To assess scaling behavior, quantum simulators were tested for $l_{\text{grid}}\in\{2,3,4\}$ while classical optimizers were extended to $l_{\text{grid}}\in\{3,\ldots,10\}$.

Justification for 36 being a sufficient sample size can be found in \cref{appen:Justi}.

\subsection{Test model}\label{sec:TM}

To test these different algorithms, we will use the following parameter values (the Mosetti benchmark case is from Ref. \cite{MOSETTI1994105}),
\begin{align*}
    D &= \text{Mosetti second benchmark case}\\
    x &= 1,\text{ }r = 1.5, \text{ }m = 4,\text{ } \xi = 0.
\end{align*}
We note that these parameters are chosen for mathematical simplicity in this initial test work and do not relate to real turbines. As a result, our toy model power output should be seen as a form of cost. $m$ is the maximum number of turbines allowed, and $\xi$ is the minimum distance allowed between turbines. In this work, we have set $\xi$ to zero, meaning turbines may be on adjacent sites. Quantum simulation constraints limit our study to an $l_{\text{grid}}=4$ case (16 turbine positions), which directly corresponds to a 16-qubit system requirement.

Mosetti's second benchmark case is defined as:
\begin{align}
    D &= \{\{10k,12ms^{-1},1/36\}\}_{k=0}^{36}\\
    &= \{\{ 0,12ms^{-1},0.028\},...,\{ 350,12ms^{-1},0.028\}\}
\end{align}

\subsection{Computational details of simulations}

The quantum simulations were conducted using IBM's \texttt{qiskit:0.39.2}
package for Python~\cite{Qiskit}. 

Both PO and COBYLA were utilized via \texttt{scipy.optimize.minimize}
- employing the methods \texttt{`powell'} and \texttt{`COBYLA'}
respectively as part of the \texttt{scipy} library~\cite{doCouto2013}. BO was coded up for this work, using a heavily modified version of the code used in \cite{BOFS}, where we no longer use \texttt{sklearn.gaussian\_process.GaussianProcess-Regressor}.
This had problems when using the periodic kernel, and so a self-made algorithm was required.

SA was implemented using the package \texttt{pyqubo} \cite{zaman2021pyqubo}.  This library has a built-in annealer: \texttt{neal.SimulatedAnnealingSampler()}. Once the system has been
turned into a QUBO problem via \texttt{model.to\_qubo()}, \texttt{pyqubo} has a wrapper for the D-Wave Ocean Software Development Kit (SDK), enabling seamless execution on D-Wave systems.

Gurobi was implemented using \texttt{gurobipy}~\cite{gurobis}, with an academic license obtained via the University of Plymouth.

As well as the previously mentioned methods, we also investigated
\texttt{qiskit}'s built-in VQE QUBO solver. This uses their \texttt{SamplingVQE} function, and then performs a \texttt{MinimumEigenOptimizer} over it to find the
groundstate. The ansatz used is the same as the one described in
\cref{sec:Q2VQE}. This method acts as a control, as it is the VQE
method with no statistical noise present. We use COBYLA as the
classical optimization routine for this method. Details about this
method can be found in Ref.~\cite{qiskitCVaR}.

All of the simulations for the results provided below were run using
the computational facilities of the High Performance Computing Centre located at the University of Plymouth. The Central Processing Units (CPUs) were dual-node Intel Xeon E5-2683v4, each with 16 cores, at the time of simulation. Task farming was used to run the simulations using all the cores.

\section{Results}

\subsection{How solutions are selected}

For both the Gurobi solver and simulated annealing, the selection of the solution is trivial; it will be the final binary string found by the algorithm.

However, for the VQE, this is less obvious. Once the optimization process has finished, we are given a set of parameters that correspond to the quantum circuit gates. We then input these values back into the circuit and measure in the $\sigma_3$ basis on all qubits; this produces a distribution over all possible basis states. From this distribution, we then take the state which has the highest probability and only has $m$ $|1\rangle$ terms.  We only accepted solutions that have exactly the correct number of turbines, which is four for our test case.

\subsection{Discussion about the degeneracy of solutions}

As for many other QUBO problems, there are many degenerate optimal solutions. For the case $l_{grid}= 4$, there are 79 unique optimal solutions out of 1820 binary vectors that meet the constraints and 65536 total possible binary strings of length 16. A list of all optimal solutions can be found in \cref{appen:Degen}. Each of these optimal solutions has a power output of 2304.0 kW.

Some of these degeneracies are due to the fact that there are optimal
solutions for the conditions in smaller system sizes. The smaller system solutions can then be embedded in the larger systems. No solutions exist with output $2304.0$ kW for $l_{grid} \leq 2$.

\begin{figure}[H]
  \centering
    \includegraphics[width=0.4\linewidth]{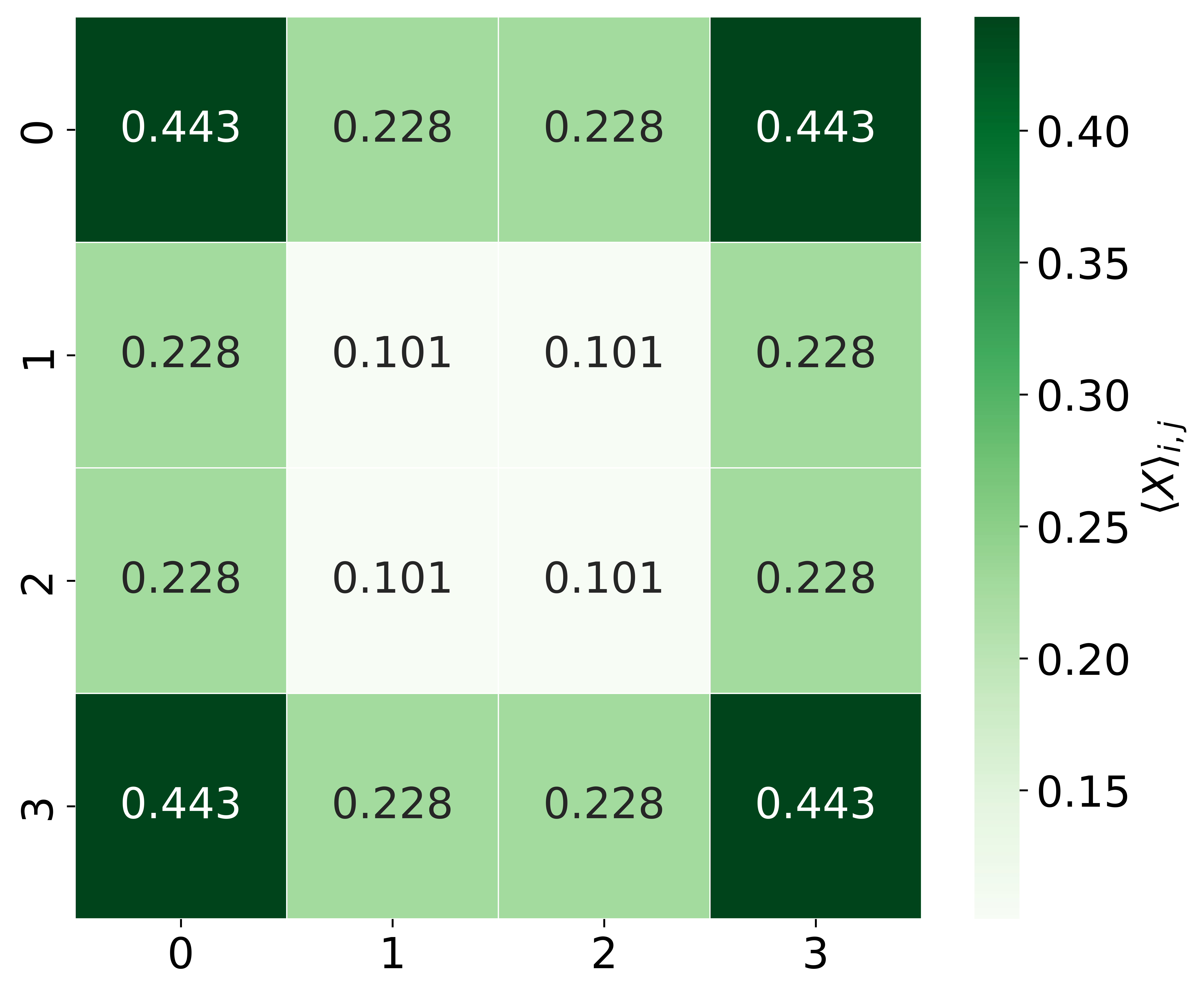}
    \caption{Heatmap of $\langle X\rangle$ values for the degenerate solutions with maximum power. All the values sum to 4 (the number of turbines).}
    \label{fig:solheat}
\end{figure}

An alternative approach to analyzing optimal solution structures involves examining average turbine placements. For a given solution $X_*^i$, represented as a binary matrix encoding turbine positions on the grid, we compute the mean location across all optimal configurations. We map from solutions to grids by using the labelling as in \cref{fig:gridcord}. We then calculate the average placement as:
\begin{equation}
    \langle X \rangle = \frac{1}{\text{\#solutions}}\sum_i X_*^i
\end{equation}
The results from this can be seen in \cref{fig:solheat}. Note here
that the sum of these should equal four, as this is the maximum number
of turbines; some rounding errors may occur due to limited precision in the printed values. From this, we see that there are more optimal solutions with turbines in the outer corners of the grid. This makes sense physically with the wind regime used, as it maximizes the distance between any two turbines and thus minimizes any possible wake effects.

\subsection{Discussion of the results }

Our simulations showed that PO achieved reasonable computation times for $l_{\text{grid}} \in \{2,3\}$, but failed to converge for $l_{\text{grid}}=4$ due to excessive shot counts and function evaluations. Consequently, we present results only for the surrogate-based methods (COBYLA and BO) and the two classical algorithms.

\begin{figure}[ht]
    \centering
    \includegraphics[width=0.8\textwidth]{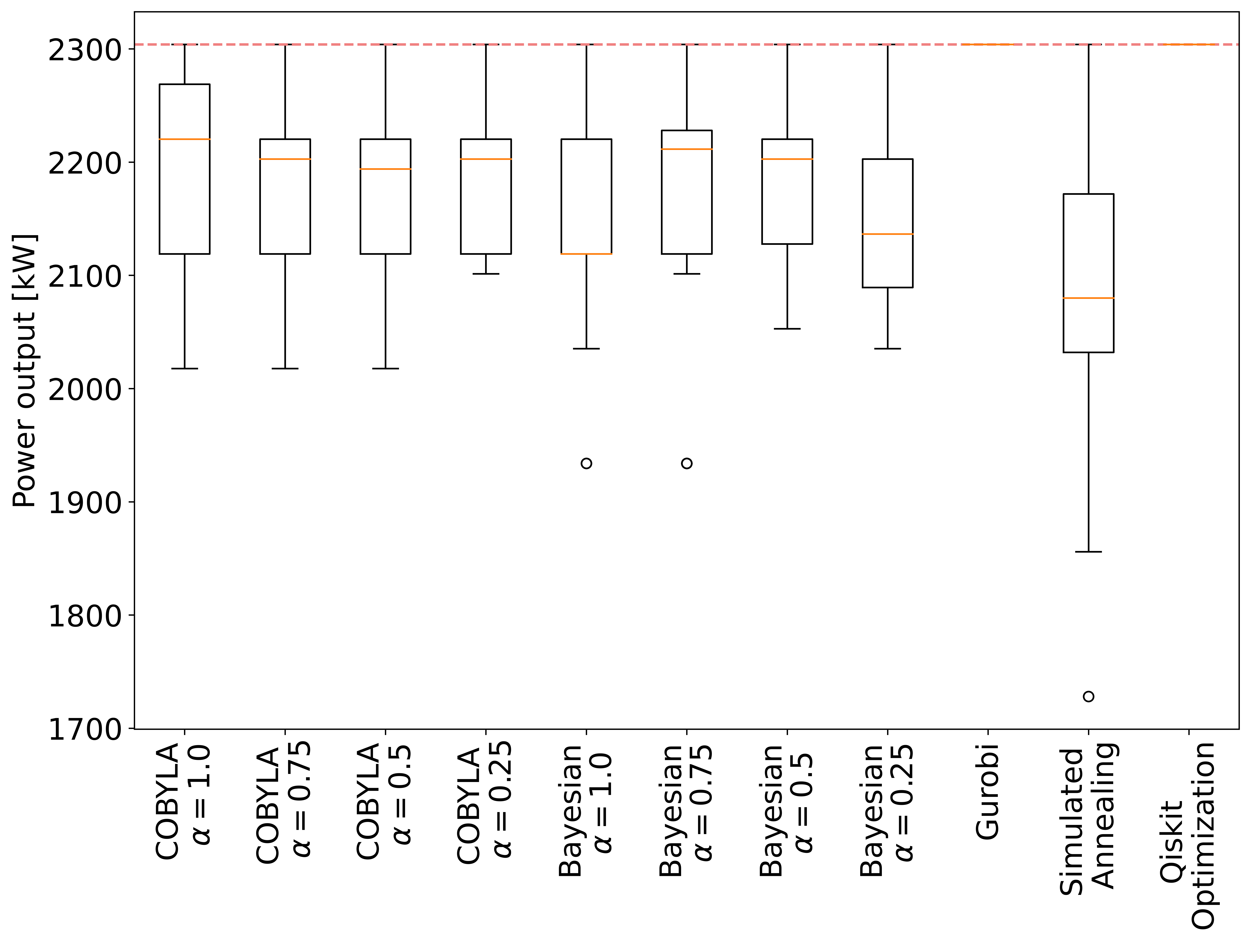}
    \caption{Each value along the $x$-axis represents the different
      methods. The box plots show the power obtained from running the
      problems 36 times. The red dashed line is the optimal solution
      power output, found through an exhaustive list search. The Gurobi
      method and Qiskit Optimization may look empty: the optimal solution was found every time. The circle points that are
      not within the boxes are outlier points.}
    \label{fig:EnergyBox}
\end{figure}

\cref{fig:EnergyBox} reports the results for the power output from
the different algorithms investigated for the $l_{grid} = 4$ case. Each method was run 36 times; the box plots show the spread of these results. The $x$-axis shows the names of the different methods, and the number under the name of the VQE method is the CVaR $\alpha$ value. 

From \cref{fig:EnergyBox} we can draw our conclusions. The Gurobi method consistently reaches the optimal solution,  outperforming all other methods. This is also true for the noiseless COBYLA-VQE method from \texttt{qiskit}. The VQE variants yield comparable solution quality both among themselves and relative to SA.

We see that each of the COBYLA-CVaR methods is capable of achieving the optimal solution. We also observe that using $\alpha<1$ has minimal impact on the quality of results. We see that the Bayesian-CVaR has very comparable results. 

\cref{tab:percents} shows the average power output of each of the methods as a percentage of the optimal power output. We see that each of these methods performs very well on average, with the lowest average output being $93.3 \%$ of the maximum.

\begin{table}[H]
    \tablesize{\footnotesize}
    \centering
    \caption{Average output power from solutions of different VQE-based methods as a percentage of the optimal power output.}
    \label{tab:percents}
    \newcolumntype{C}{>{\centering\arraybackslash}X}
    \begin{tabularx}{0.6\textwidth}{CC}  
        \toprule
        \textbf{Method} & \textbf{Average power output (\% of optimal)} \\
        \midrule
        COBYLA-1.00 & 95.5 \\
        COBYLA-0.75 & 94.7 \\
        COBYLA-0.50 & 94.5 \\
        COBYLA-0.25 & 95.2 \\
        Bayes-1.00 & 93.3 \\
        Bayes-0.75 & 94.2 \\
        Bayes-0.50 & 94.6 \\
        Bayes-0.25 & 93.4 \\
        \bottomrule
    \end{tabularx}
\end{table}

\subsection{Timing the results}\label{sec:timing}

Beyond solution quality, we assess computational efficiency by measuring algorithm runtime. Using Python's \texttt{time} module, we recorded end-to-end execution times, with complete results presented in \cref{tab:timings}.

\begin{table}[H]
    \tablesize{\footnotesize}
    \centering
    \newcolumntype{C}{>{\centering\arraybackslash}X}
    \begin{tabularx}{\textwidth}{CCCCC}
        \toprule
        \boldmath$l_{grid}$ & \textbf{Gurobi (s)} & \textbf{Simulated Annealing (s)} & \textbf{COBYLA-CVaR (s)} & \textbf{BO-CVaR (s)} \\
        \midrule
        2 & -- & -- & 19.74 & 250.11 \\
        3 & 0.09 & 0.04 & 561.78 & 5741.43 \\
        4 & 1.11 & 0.85 & 11546.10 & 55454.77 \\
        5 & 3.91 & 3.31 & -- & -- \\
        6 & 14.51 & 11.49 & -- & -- \\
        7 & 107.11 & 106.62 & -- & -- \\
        8 & 374.72 & 377.49 & -- & -- \\
        9 & 983.27 & 985.57 & -- & -- \\
        10 & 2240.78 & 2220.44 & -- & -- \\
        \bottomrule
    \end{tabularx}
    \caption{Average time in seconds taken for the different optimization algorithms. Note: All timings are averaged results. Missing entries (--) indicate the method was not evaluated at this grid size.}\label{tab:timings}
\end{table}

These results demonstrate that the quantum simulator exhibits significantly longer runtimes than classical algorithms at fixed system sizes. More importantly, analyzing how runtime scales with increasing system size reveals each algorithm's empirical computational complexity, which gives us a good metric as quantum resources grow in size. We can study this by looking at the regression coefficients of LogLog plots of the timings.

\begin{figure}[H]
    \centering
    \includegraphics[width=0.7\textwidth]{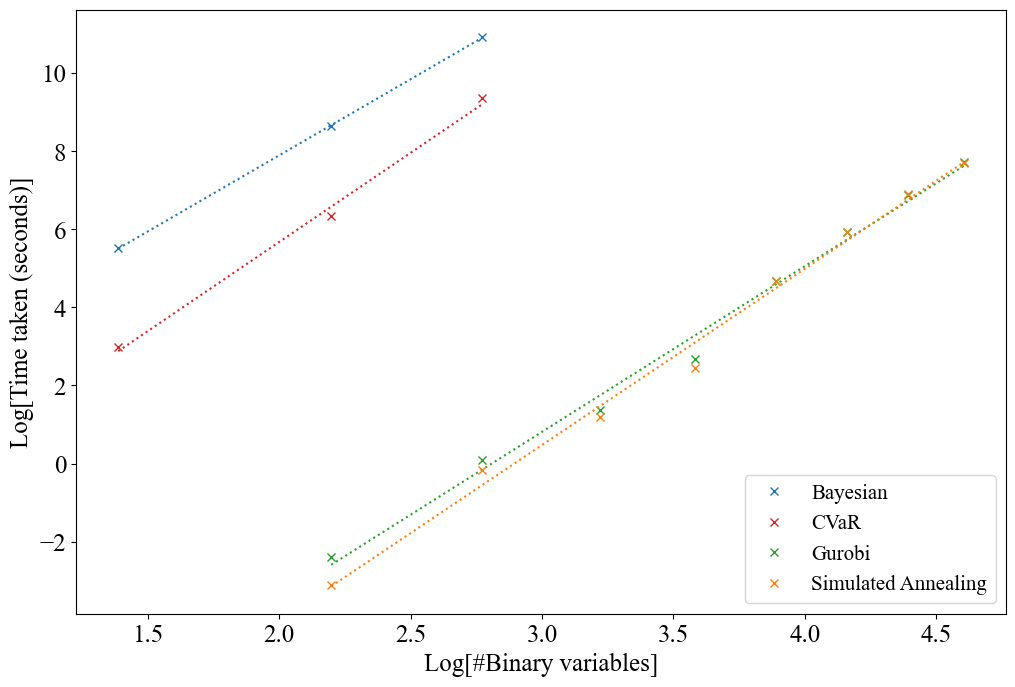}
    \caption{LogLog plots of the time taken for the different methods over different system sizes.}
    \label{fig:logtimings}
\end{figure}

The time to solve the QUBO problem is expected to depend on the system size exponentially. While the worst-case complexity is exponential, our results show power-law scaling, due to the simplicity of the scaling problem and limited data. As we only have three data points, we parameterize the time to solution as a polynomial with exponent $\alpha$ as
\begin{equation*}
    \text{Time to solve} \propto (V)^{\alpha},
\end{equation*}
where $V=l_{grid}^2$ is the system volume. \cref{fig:logtimings} shows the Log of time taken to the solution versus the Log of the system volume. The power-law fit coefficients can be found in \cref{tab:tcoefs}. The results from all four algorithms in \cref{fig:logtimings} have similar slopes.
\begin{table}[H]
    \tablesize{\footnotesize}
    \centering
    \caption{Slope and intercept from linear regression of Log(time) vs. Log(number of binary variables)
for the different algorithms.}
    \label{tab:tcoefs}
    \newcolumntype{C}{>{\centering\arraybackslash}X}
    \begin{tabularx}{0.7\textwidth}{lCC}  
        \toprule
        \textbf{Method} & \textbf{Gradient} & \textbf{Intercept} \\
        \midrule
        Gurobi & 4.24 & -5.18 \\
        Simulated Annealing & 4.51 & -5.67 \\
        CVaR-VQE using COBYLA & 4.57 & -1.50 \\
        CVaR-VQE using BO & 3.89 & 0.04 \\
        \bottomrule
    \end{tabularx}
\end{table}

We find that: Gurobi $\sim O(V^{4.24})$, SA $\sim O(V^{4.51})$, COBYLA $\sim O(V^{4.57})$ and Bayesian $\sim O(V^{3.89})$. The excellent LogLog linear fits support this approximation for our data range. This result is particularly noteworthy because while Gurobi and SA represent actual algorithm implementations, the quantum approach runs on a classical simulator. We anticipate that true quantum hardware would significantly reduce these computation times.

\begin{figure}[H]
    \centering
    \begin{minipage}{0.45\textwidth}
    \centering
        \includegraphics[width=\linewidth]{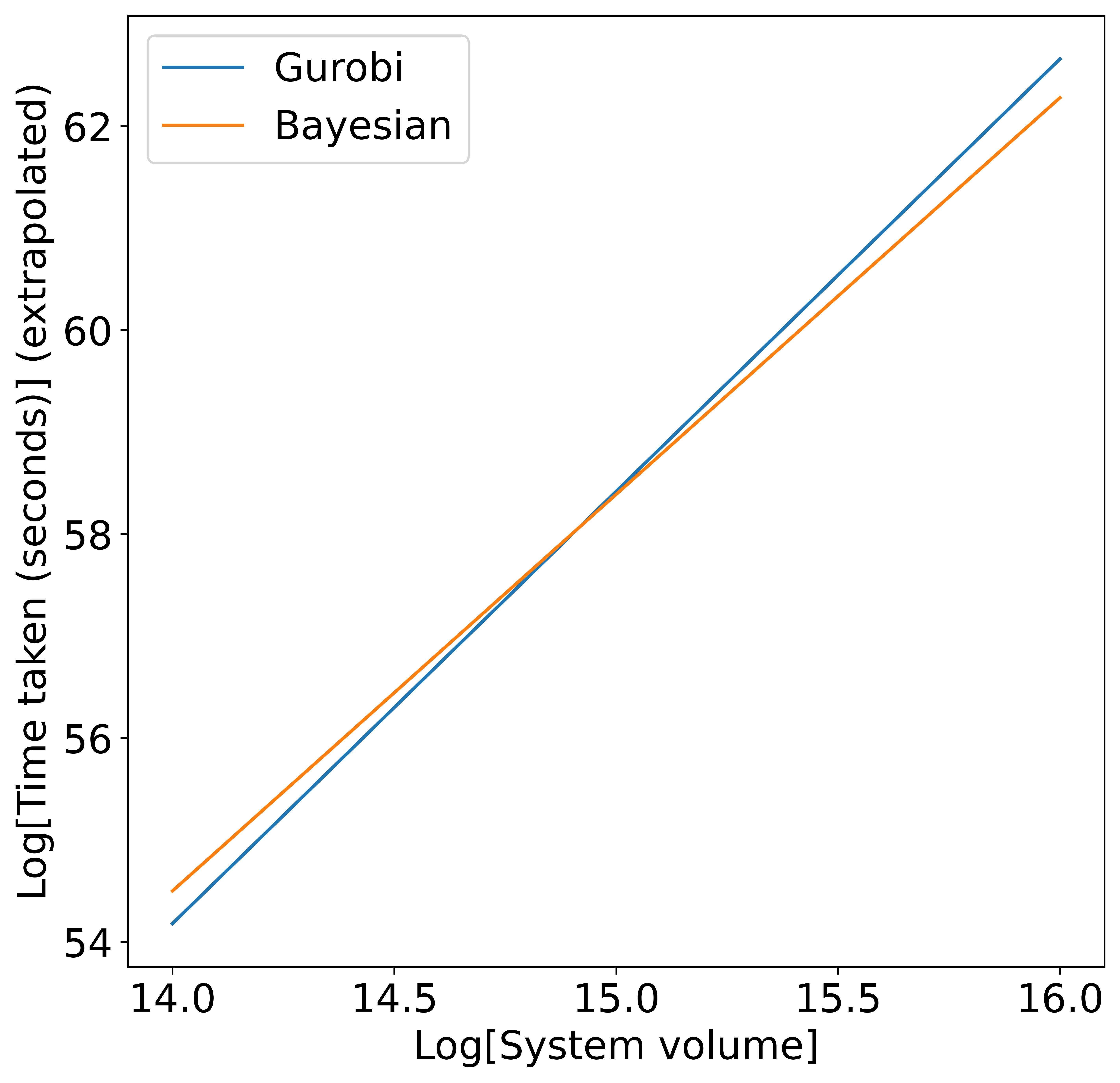}
        \caption{Extrapolation of time taken for Gurobi and Bayesian Optimization to find the point at which Bayesian would be faster.}
  \label{fig:BAYESGurobi}
    \end{minipage}\hfill
    \begin{minipage}{0.45\textwidth}
        \centering
        \includegraphics[width=\linewidth]{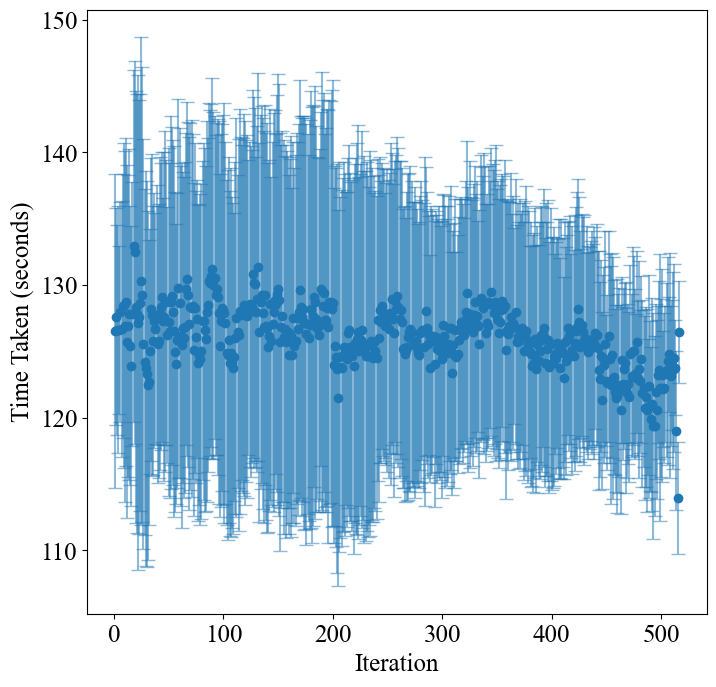}
        \caption{\corr{Time taken per iteration when using COBYLA applied to our WFLO problem. Each iteration takes the same amount of time as the previous.}}
        \label{fig:timePerIter_COBYLA}
    \end{minipage}
    
\end{figure}

We note that BO, run on the simulator, scales better than all the other methods, having the lowest polynomial scaling; this does come with the caveat that the pre-factor is much larger than the other systems. We can look at this more closely by searching for the intersection point of the Bayesian and Gurobi timing lines, shown in \cref{fig:BAYESGurobi}. We can estimate that the simulated Bayesian-CVaR method would be faster than the Gurobi method for a system volume of $\approx 10^{15}$, which is a square grid, with $l_\text{grid}\approx 31622776$. This scaling extrapolation shows that even when simulating, there may be a point at which this algorithm is faster. However, more data at larger scales is required to make a concrete claim, to increase the accuracy of our power-law scaling approximation. We emphasize that these scaling laws are empirical observations, not theoretical complexity predictions.
\begin{figure}[H]
    \centering
    \begin{minipage}{0.45\textwidth}
        \centering
        \includegraphics[width=\linewidth]{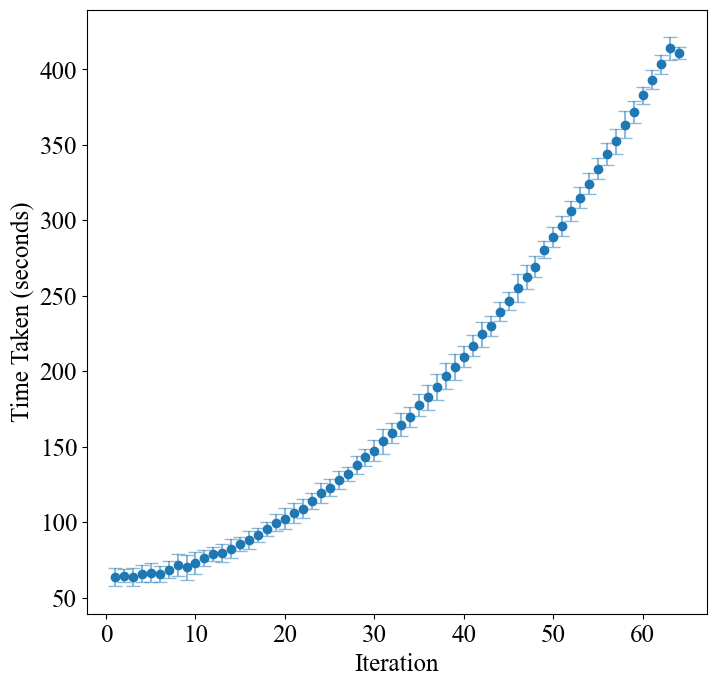}
        \caption{\corr{Time taken per iteration when using BO applied to our WFLO problem. The time taken seems to scale exponentially.}}
        \label{fig:timePerIter_Bayes}
    \end{minipage}\hfill
    \begin{minipage}{0.45\textwidth}
        \centering
        \includegraphics[width=\linewidth]{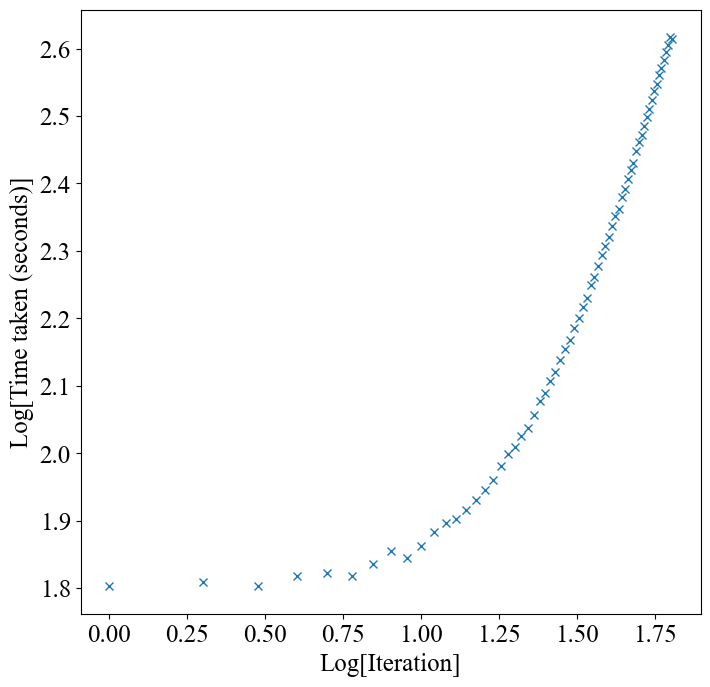}
        \caption{\corr{LogLog plot of time taken per iteration when using BO applied to our WFLO problem. The time taken seems to scale exponentially.}}
        \label{fig:timePerIter_Bayesian_LL}
    \end{minipage}
\end{figure}
Another important consideration when comparing COBYLA and BO in this setting is the time taken per iteration. \cref{fig:timePerIter_COBYLA,fig:timePerIter_Bayes} show the iteration timings for runs of our WFLO problem with $l_{\text{grid}} = 4$ (each of the data points shows samples from 64 runs, where the points are the average and the error bars are the standard deviation of these samples), for COBYLA and BO respectively. For COBYLA, the time per iteration remains approximately constant across the optimization, indicating no significant scaling. In contrast, for BO, the time per iteration increases, appearing to scale exponentially, as highlighted in the LogLog plot in \cref{fig:timePerIter_Bayesian_LL}. Despite this exponential growth in time per iteration, the overall time to solution is \emph{better}, as the problem scales, compared to COBYLA, as demonstrated in \cref{fig:logtimings}.

\subsection{Conclusions}
We successfully demonstrated the novel application of quantum circuit methods to the WFLO problem, establishing this as a promising new approach for finding optimal turbine configurations. Our quantum-classical hybrid method shows particular potential for future quantum hardware implementations. The VQE-based methods demonstrate viability for identifying optimal or near-optimal solutions while exhibiting favorable time complexity scaling, provided a sufficient number of measurements have been performed. The Gurobi optimizer always finds the optimal solution, outperforming the noisy VQE-based method.  This is not surprising, as Gurobi has been developed to specifically solve linear and quadratic programming problems, of which QUBO is a special case.  

We have only investigated small system sizes due to computational constraints in the simulator. Numerous benchmarking studies have evaluated classical algorithms for WFLO problems using high-resolution grids, typically at least 100 points. This would require 100 qubits in our VQE-based approach, and possibly many more when error correction is used.

Interestingly, we observe comparable scaling exponents in runtime versus system size between Gurobi and the simulated VQE approach. From our results, the performance of the VQE method critically depends on the performance of the optimizer that runs on the classical computer.
 
The development of the VQE is an active area of research. For example, \cite{kungurtsev2023iteration} has studied the iteration complexity of Variational Quantum Algorithms in a noisy environment. A generalization of CVaR-VQE has been proposed, called the Filter-VQE (F-VQE). F-VQE uses a technique based on filtering operators to achieve faster and more reliable convergence to the optimal solution \cite{Amaro_2022}. When dealing with high gate count variational circuits, exponentially vanishing gradients, or \textit{barren plateaus} \cite{Cerezo_2021}, have been found. When scaling to the system sizes discussed in \cref{sec:timing}, barren plateaus will become increasingly prevalent due to the growing number of circuit parameters. Recent work \cite{mastropietro2023flemingviot} addresses this challenge through parallel optimization routines (particles) that dynamically relocate when encountering vanishing gradients or noise-dominated measurements.

We are encouraged by the results for the solution of the WFLO obtained on the quantum computing simulator. Next, we will study the resilience of the implemented methods against quantum errors on the simulators, before using a quantum computer. It will be interesting to compare the performance on circuit-based quantum computers to that of adiabatic quantum systems. 

Ultimately, as quantum computers become larger and more reliable, we hope that the methods we have begun to develop here will help wind farms produce more power, accelerating the transition to Net Zero.

\subsection{Acknowledgments}
We thank Julian Stander for the discussions about Bayesian optimization. 
The simulations for this work were carried out using the computational facilities of the
High Performance Computing Centre located at the University of Plymouth -- \url{https://www.plymouth.ac.uk/about-us/university-structure/faculties/science-engineering/hpc}.

\subsection{Funding statement}
The authors received no specific funding for this study.

\subsection{Authorship contribution statement}
James Hancock: Writing – review \& editing, Writing – original draft, Visualization, Validation, Software, Project administration,
Methodology, Investigation, Formal analysis, Data curation, Conceptualization. Matthew Craven: Writing – review \& editing, Writing –
original draft, Visualization, Validation, Supervision, Project administration, Methodology, Investigation,  Formal analysis, Conceptualization. Craig McNeile: Writing – review \& editing,
Visualization, Validation, Supervision, Project
administration, Methodology, Investigation, Conceptualization.
Davide Vadacchino: Writing – review \& editing, Supervision, Project
administration, Methodology, Investigation, Conceptualization.

\subsection{Availability of data and materials}
The data and simulation code that support the findings of this study are openly available at GitHub: \url{https://github.com/jamesshancock/WindfarmLayoutOptimization}.

\subsection{Ethics approval}
Not applicable.

\subsection{Conflicts of interest}
The authors declare no conflicts of interest to report regarding the present study.


\begin{thebibliography}{10}

\bibitem{veers2019grand}
Veers P, Dykes K, Lantz E, Barth S, Bottasso CL, Carlson O, et~al.
\newblock Grand challenges in the science of wind energy.
\newblock Science. 2019;366(6464):eaau2027.
\newblock Available from: \url{https://doi.org/10.1126/science.aau2027}.

\bibitem{giani2021quantum}
Giani A, Eldredge Z.
\newblock Quantum computing opportunities in renewable energy.
\newblock SN Computer Science. 2021;2(5):393.
\newblock Available from: \url{https://doi.org/10.1038/s43588-021-00032-z}.

\bibitem{ajagekar2022quantum}
Ajagekar A, You F.
\newblock Quantum computing and quantum artificial intelligence for renewable
  and sustainable energy: A emerging prospect towards climate neutrality.
\newblock Renewable and Sustainable Energy Reviews. 2022;165:112493.
\newblock Available from: \url{https://doi.org/10.1016/j.rser.2022.112493}.

\bibitem{abbas2023quantum}
Abbas A, Ambainis A, Augustino B, B{\"a}rtschi A, Buhrman H, Coffrin C, et~al.
\newblock Quantum Optimization: Potential, Challenges, and the Path Forward.
\newblock arXiv preprint arXiv:231202279. 2023.
\newblock Available from: \url{https:/doi.org/10.2172/2229681}.

\bibitem{MOSETTI1994105}
Mosetti G, Poloni C, Diviacco B.
\newblock Optimization of wind turbine positioning in large windfarms by means
  of a genetic algorithm.
\newblock Journal of Wind Engineering and Industrial Aerodynamics.
  1994;51(1):105-16.
\newblock Available from: \url{https://doi.org/10.1016/0167-6105(94)90080-9}.

\bibitem{manikowski2021multi}
Manikowski PL, Walker DJ, Craven MJ.
\newblock Multi-Objective Optimisation of the Benchmark Wind Farm Layout
  Problem.
\newblock J Mar Sci Eng. 2021;9(12):1376.
\newblock Available from: \url{https://doi.org/10.3390/jmse9121376}.

\bibitem{senderovich2022exploiting}
Senderovich A, Zhang J, Cohen E, Beck JC.
\newblock Exploiting Hardware and Software Advances for Quadratic Models of
  Wind Farm Layout Optimization.
\newblock IEEE Access. 2022;10:78044-55.
\newblock Available from: \url{https://doi.org/10.1109/ACCESS.2022.3193143}.

\bibitem{QAoverview}
Rajak A, Suzuki S, Dutta A, Chakrabarti BK.
\newblock Quantum annealing: An overview.
\newblock Philosophical Transactions of the Royal Society A: Mathematical,
  Physical and Engineering Sciences. 2023;381(2241):20210417.
\newblock Available from: \url{https://doi.org/10.1098/rsta.2021.0417}.

\bibitem{DWAd}
McGeoch C, Farré P.
\newblock Advantage processor overview.
\newblock D-Wave Systems. 2022.
\newblock Has no DOI available.
\newblock Available from:
  \url{https://www.dwavequantum.com/resources/white-paper/the-d-wave-advantage-system-an-overview/}.

\bibitem{DWAd2}
McGeoch C, Farré P, Boothby K.
\newblock D-WAVE Advantage2 prototype.
\newblock D-Wave Systems. 2022.
\newblock Available from:
  \url{https://www.dwavequantum.com/media/eixhdtpa/14-1063a-a_the_d-wave_advantage2_prototype-4.pdf}.

\bibitem{matsubara2020digital}
Matsubara S, Takatsu M, Miyazawa T, Shibasaki T, Watanabe Y, Takemoto K, et~al.
\newblock Digital Annealer for High-Speed Solving of Combinatorial Optimization
  Problems and Its Applications.
\newblock In: Proc. 25th Asia South Pacific Design Autom. Conf. (ASP-DAC).
  IEEE; 2020. p. 667-72.
\newblock Available from:
  \url{https://doi.org/10.1109/ASP-DAC47756.2020.9045100}.

\bibitem{Qiskit}
{Qiskit contributors}. Qiskit: An Open-source Framework for Quantum Computing;
  2023.
\newblock Available from: \url{https://doi.org/10.5281/zenodo.2573505}.

\bibitem{preskill2018quantum}
Preskill J.
\newblock Quantum computing in the {NISQ} era and beyond.
\newblock Quantum. 2018;2:79.
\newblock Available from: \url{https://doi.org/10.22331/q-2018-08-06-79}.

\bibitem{abbas2024challenges}
Abbas A, Ambainis A, Augustino B, B{\"a}rtschi A, Buhrman H, Coffrin C, et~al.
\newblock Challenges and opportunities in quantum optimization.
\newblock Nature Reviews Physics. 2024:1-18.
\newblock Available from: \url{https://doi.org/10.1038/s42254-024-00770-9}.

\bibitem{zahedinejad2017combinatorial}
Zahedinejad E, Zaribafiyan A.
\newblock Combinatorial optimization on gate model quantum computers: A survey.
\newblock arXiv preprint. 2017;arXiv:1708.05294.
\newblock Available from: \url{https://doi.org/10.48550/arXiv.1708.05294}.

\bibitem{albash2018demonstration}
Albash T, Lidar DA.
\newblock Demonstration of a scaling advantage for a quantum annealer over
  simulated annealing.
\newblock Phys Rev X. 2018;8(3):31016.
\newblock Available from: \url{https://doi.org/10.1103/PhysRevX.8.031016}.

\bibitem{pearson2019analog}
Pearson A, Mishra A, Hen I, Lidar DA.
\newblock Analog errors in quantum annealing: doom and hope.
\newblock npj Quantum Information. 2019;5(1):107.
\newblock Available from: \url{https://doi.org/10.1038/s41534-019-0210-7}.

\bibitem{yaacoby2022comparison}
Yaacoby R, Schaar N, Kellerhals L, Raz O, Hermelin D, Pugatch R.
\newblock Comparison between a quantum annealer and a classical approximation
  algorithm for computing the ground state of an Ising spin glass.
\newblock Phys Rev E. 2022;105(3):035305.
\newblock Available from: \url{https://doi.org/10.1103/PhysRevE.105.035305}.

\bibitem{martin2015unraveling}
Martin-Mayor V, Hen I.
\newblock Unraveling Quantum Annealers Using Classical Hardness.
\newblock Scientific Reports. 2015;5(1):15324.
\newblock Available from: \url{https://doi.org/10.1038/srep15324}.

\bibitem{SSproceedings}
Zhang P, Romero D, Beck JC, Amon C.
\newblock Solving Wind Farm Layout Optimization with Mixed Integer Programming
  and Constraint Programming.
\newblock In: Integration of AI and OR Techniques in Constraint Programming for
  Combinatorial Optimization Problems. Berlin, Heidelberg: Springer; 2013. p.
  284-99.
\newblock Available from: \url{https://doi.org/10.1007/978-3-642-38171-3_19}.

\bibitem{Donovan2005WindFO}
S D.
\newblock An improved mixed integer programming model for wind farm layout
  optimisation.
\newblock In: Proc. 41st Annu. Conf. Oper. Res. Soc. N. Z.; 2006. p. 143-52.
\newblock Has no DOI available.
\newblock Available from:
  \url{https://www.researchgate.net/publication/242744637}.

\bibitem{QUBOsurvey}
Kochenberger G, Hao JK, Glover F, Lewis M, Lü Z, Wang H, et~al.
\newblock The unconstrained binary quadratic programming problem: A survey.
\newblock Journal of Combinatorial Optimization. 2014;28(1):58-81.
\newblock Available from: \url{https://doi.org/10.1007/s10878-014-9734-0}.

\bibitem{khanali2018optimizing}
Khanali M, Ahmadzadegan S, Omid M, Keyhani~Nasab F, Chau KW.
\newblock Optimizing Layout of Wind Farm Turbines Using Genetic Algorithms in
  Tehran Province, Iran.
\newblock Int J Energy Environ Eng. 2018;9:399-411.
\newblock Available from: \url{https://doi.org/10.1007/s40095-018-0279-3}.

\bibitem{gurobis}
{Gurobi Optimization LLC}. Gurobi Optimizer Reference Manual; 2024.
\newblock Accessed: 27-06-2023.
\newblock Available from:
  \url{https://docs.gurobi.com/projects/optimizer/en/current/index.html}.

\bibitem{cseker2022digital}
{\c{S}}eker O, Tanoumand N, Bodur M.
\newblock Digital Annealer for quadratic unconstrained binary optimization: A
  comparative performance analysis.
\newblock Appl Soft Comput. 2022;127:109367.
\newblock Available from: \url{https://doi.org/10.1016/j.asoc.2022.109367}.

\bibitem{nakano2022diverse}
Nakano K, Takafuji D, Ito Y, Yazane T, Yano J, Ozaki S, et~al.
\newblock Diverse Adaptive Bulk Search: a Framework for Solving QUBO Problems
  on Multiple GPUs.
\newblock In: Proc. 2023 IEEE Int. Parallel Distrib. Process. Symp. Workshops
  (IPDPSW). IEEE; 2023. p. 314-25.
\newblock Available from: \url{https://doi.org/10.1109/IPDPSW59300.2023.00060}.

\bibitem{kirkpatrick1983optimization}
Kirkpatrick S, Gelatt CD, Vecchi MP.
\newblock Optimization by simulated annealing.
\newblock Science. 1983;220(4598):671-80.
\newblock Available from: \url{https://doi.org/10.1126/science.220.4598.671}.

\bibitem{SATSP}
Wang Z, Geng X, Shao Z.
\newblock An Effective Simulated Annealing Algorithm for Solving the Traveling
  Salesman Problem.
\newblock Journal of Computational and Theoretical Nanoscience.
  2009;6(7):1680-6.
\newblock Available from: \url{https://doi.org/10.1166/jctn.2009.1230}.

\bibitem{SAMLA}
Rodriguez-Tello E, Hao JK, Torres-Jimenez J.
\newblock An effective two-stage simulated annealing algorithm for the minimum
  linear arrangement problem.
\newblock Computers \& Operations Research. 2008;35(10):3331-46.
\newblock Available from: \url{https://doi.org/10.1016/j.cor.2007.03.001}.

\bibitem{SAPP}
Tole K, Moqa R, Zheng J, He K.
\newblock A simulated annealing approach for the circle bin packing problem
  with rectangular items.
\newblock Computers \& Industrial Engineering. 2023;176:109004.
\newblock Available from: \url{https://doi.org/10.1016/j.cie.2023.109004}.

\bibitem{VQEorig}
Peruzzo A, McClean J, Shadbolt P, Yung MH, Zhou XQ, Love PJ, et~al.
\newblock A Variational Eigenvalue Solver on a Photonic Quantum Processor.
\newblock Nat Commun. 2014;5:4213.
\newblock Available from: \url{https://doi.org/10.1038/ncomms5213}.

\bibitem{VQE}
Tilly J, Chen H, Cao S, Picozzi D, Setia K, Li Y, et~al.
\newblock The variational quantum eigensolver: A review of methods and best
  practices.
\newblock Phys Rep. 2022;986:1-128.
\newblock Available from: \url{https://doi.org/10.1016/j.physrep.2022.08.003}.

\bibitem{COBYLA1}
Liu X, Angone A, Shaydulin R, Safro I, Alexeev Y, Cincio L.
\newblock Layer VQE: A Variational Approach for Combinatorial Optimization on
  Noisy Quantum Computers.
\newblock IEEE Trans Quantum Eng. 2022 January;3:1-20.
\newblock Available from: \url{https://doi.org/10.1109/TQE.2021.3140190}.

\bibitem{COBYLA2}
Schwägerl T, Issever C, Jansen K, Khoo T, Kühn S, Tüysüz C, et~al.
\newblock Particle track reconstruction with noisy intermediate-scale quantum
  computers.
\newblock arXiv preprint. 2023.
\newblock Available from: \url{https://doi.org/10.48550/arXiv.2303.13249}.

\bibitem{McClean:2015vup}
McClean JR, Romero J, Babbush R, Aspuru-Guzik A.
\newblock The theory of variational hybrid quantum-classical algorithms.
\newblock New Journal of Physics. 2016;18(2):023023.
\newblock Available from: \url{https://doi.org/10.1088/1367-2630/18/2/023023}.

\bibitem{Klco:2018kyo}
Klco N, Dumitrescu EF, McCaskey AJ, Morris TD, Pooser RC, Sanz M, et~al.
\newblock Quantum-classical computation of Schwinger model dynamics using
  quantum computers.
\newblock Phys Rev A. 2018 Sep;98(3):032331.
\newblock Available from: \url{https://doi.org/10.1103/PhysRevA.98.032331}.

\bibitem{PBh}
Snoek J, Larochelle H, Adams RP.
\newblock Practical Bayesian Optimization of Machine Learning Algorithms.
\newblock In: Pereira F, Burges CJ, Bottou L, Weinberger KQ, editors. Advances
  in Neural Information Processing Systems. vol.~25. Red Hook, NY, United
  States: Curran Associates, Inc.; 2012. Available from:
  \url{https://dl.acm.org/doi/10.5555/2999325.2999464}.

\bibitem{BOa}
Nguyen V.
\newblock Bayesian Optimization for Accelerating Hyper-Parameter Tuning.
\newblock In: 2019 IEEE Second International Conference on Artificial
  Intelligence and Knowledge Engineering (AIKE). IEEE; 2019. p. 302-5.
\newblock Available from: \url{https://doi.org/10.1109/AIKE.2019.00060}.

\bibitem{AfML}
Bergstra J, Bardenet R, Bengio Y, Kégl B.
\newblock Algorithms for Hyper-Parameter Optimization.
\newblock Advances in Neural Information Processing Systems. 2011;24.
\newblock Available from: \url{https://dl.acm.org/doi/10.5555/2986459.2986743}.

\bibitem{KJ}
Iannelli G, Jansen K.
\newblock Noisy Bayesian optimization for variational quantum eigensolvers.
\newblock In: LATTICE2021. vol. 396. Trieste, Italy; 2022. p. 251.
\newblock Available from: \url{https://doi.org/10.22323/1.396.0251}.

\bibitem{Funcke_2021}
Funcke L, Hartung T, Jansen K, Kühn S, Stornati P.
\newblock Dimensional {E}xpressivity {A}nalysis of {P}arametric {Q}uantum
  {C}ircuits.
\newblock Quantum. 2021;5:422.
\newblock Available from: \url{https://doi.org/10.22331/q-2021-03-29-422}.

\bibitem{funcke2021dimensional}
Hartung T, Funcke L, Jansen K, Kühn S, Schneider M, Stornati P.
\newblock Dimensional Expressivity Analysis, best-approximation errors, and
  automated design of parametric quantum circuits.
\newblock In: LATTICE2021. Trieste, Italy; 2022. p. 575.
\newblock Available from: \url{https://doi.org/10.22323/1.396.0575}.

\bibitem{CVaR}
Barkoutsos PK, Nannicini G, Robert A, Tavernelli I, Woerner S.
\newblock Improving Variational Quantum Optimization using CVaR.
\newblock Quantum. 2020 April;4:256.
\newblock Available from: \url{https://doi.org/10.22331%2Fq-2020-04-20-256}.

\bibitem{doCouto2013}
do~Couto TG, Farias B, Diniz ACGC, de~Morais MVG.
\newblock Optimization of Wind Farm Layout Using Genetic Algorithm.
\newblock In: Proc. 10th World Congr. Struct. Multidiscip. Optim.; 2013. p.
  1-10.
\newblock Available from: \url{https://doi.org/10.1051/e3sconf/202458101024}.

\bibitem{BOFS}
Brownlee J. How to Implement Bayesian Optimization from Scratch in Python;.
\newblock Accessed: 27-06-2023.
\newblock https://machinelearningmastery.com/what-is-bayesian-optimization/.

\bibitem{zaman2021pyqubo}
Zaman M, Tanahashi K, Tanaka S.
\newblock PyQUBO: Python Library for Mapping Combinatorial Optimization
  Problems to QUBO Form.
\newblock IEEE Transactions on Computers. 2022;71(4):838-50.
\newblock Available from: \url{https://doi.org/10.1109/TC.2021.3063618}.

\bibitem{qiskitCVaR}
{Qiskit Optimization Team}. {CVaR Optimization — Qiskit Optimization
  Tutorials}; 2024.
\newblock Accessed: 2025-05-08.
\newblock Available from:
  \url{https://qiskit-community.github.io/qiskit-optimization/tutorials/08_cvar_optimization.html}.

\bibitem{kungurtsev2023iteration}
Kungurtsev V, Korpas G, Marecek J, Zhu EY.
\newblock Iteration complexity of variational quantum algorithms.
\newblock Quantum. 2024;8:1495.
\newblock Available from: \url{https://doi.org/10.22331/q-2024-10-10-1495}.

\bibitem{Amaro_2022}
Amaro D, Modica C, Rosenkranz M, Fiorentini M, Benedetti M, Lubasch M.
\newblock Filtering variational quantum algorithms for combinatorial
  optimization.
\newblock Quantum Sci Technol. 2022;7(1):15021.
\newblock Available from: \url{https://dx.doi.org/10.1088/2058-9565/ac3e54}.

\bibitem{Cerezo_2021}
Cerezo M, Sone A, Volkoff T, Cincio L, Coles PJ.
\newblock Cost function dependent barren plateaus in shallow parametrized
  quantum circuits.
\newblock Nat Commun. 2021;12(1):1791.
\newblock Available from: \url{https://doi.org/10.1038/s41467-021-21728-w}.

\bibitem{mastropietro2023flemingviot}
Mastropietro D, Korpas G, Kungurtsev V, Marecek J.
\newblock Parallel variational quantum algorithms with gradient-informed
  restart to speed up optimisation in the presence of barren plateaus.
\newblock arXiv preprint. 2024;2311.18090.
\newblock Available from: \url{https://doi.org/10.48550/arXiv.2311.18090}.

\end{thebibliography}

\appendixstart
\appendix

\section[\appendixname~\thesection]{Dimensional Expressivity Analysis}\label{appen:DEA}
Circuit expressivity analysis provides a systematic approach to evaluate and enhance system performance by quantifying the parameterized quantum circuit's capability to explore the solution space. For an objective function $F\left(\overrightarrow{\theta}\right)$, we can then perturb $\theta_k$ by $\delta\theta_k$ to produce $F\left(\overrightarrow{\theta} + \hat{e}_k\delta\theta_k\right) = f$. If we can also have $F\left(\overrightarrow{\theta} + \sum_{i\neq k}\hat{e}_i\delta\theta_i\right) = f$, $\theta_k$ is redundant. To perform this analysis, we are using the method described by Lena Funcke et al. in Ref. \cite{funcke2021dimensional}. The method is as follows:

\begin{enumerate}
    \item This can be checked by considering the real partial Jacobians $J_k$ of $C$ ($C$ is the matrix of the circuit applied to the state $|0\rangle^{\otimes q} $):
    \begin{equation*}
        J_k(\theta) = \begin{pmatrix}
        Re(\partial_1 C) ... Re(\partial_k C)\\
        Im(\partial_1 C) ... Im(\partial_k C)
        \end{pmatrix}
    \end{equation*}
    Here it is key to understand that $\partial_k C$ is itself a $2^q\times 1$ vector, and thus $J_k$ is a $2^{q+1}\times k$ matrix, where we start with $k=1$ for the first parameter.
    
    \item We then check the rank of the matrix $J_k$ for each $k$, and if adding a new parameter does not increase the rank, the parameter must be redundant. 
\end{enumerate}

To efficiently check the rank of $J_k$, we consider the matrix $S_k = J_k^*J_k$. Thus, if we check that $S_k$ is invertible, we know that all the parameters are independent (i.e., not redundant).

\section[\appendixname~\thesection]{Justification for Thirty-Six Samples} \label{appen:Justi}
When gathering results for this work, we collected 36 samples per optimization algorithm. Here, we focus specifically on COBYLA with CVaR ($\alpha = 0.25$), analyzing an extended sample set (284 runs) to validate our initial findings. The statistical similarity between the larger sample and our original 36-run dataset confirms the adequacy of the smaller sample size for drawing robust conclusions.

The initial 36 samples taken were:
\begin{verbatim}
    2136.54, 2118.96, 2202.69, 2220.27, 
    2220.27, 2220.27, 2268.84, 
    2220.27, 2286.42, 2220.27, 2136.54, 
    2118.96, 2118.96, 2220.27, 
    2101.38, 2202.69, 2220.27, 2286.42,
    2202.69, 2286.42, 2304.0, 
    2118.96, 2286.42, 2118.96, 2118.96, 
    2136.54, 2136.54, 2268.84, 
    2118.96, 2118.96, 2101.38, 2286.42,
    2220.27, 2286.42, 2202.69, 
    2118.96
\end{verbatim}
The extra 284 samples taken were:
\begin{verbatim}
    2251.25, 2202.69, 2185.11, 2185.11, 
    2136.54, 2136.54, 2286.42, 
    2118.96, 2268.84, 2304.0, 2202.69, 
    2136.54, 2220.27, 2268.84, 
    2101.38, 2118.96, 2268.84, 2220.27, 
    2220.27, 2202.69, 2202.69, 
    1933.92, 2118.96, 2202.69, 2268.84,
    2220.27, 2220.27, 2220.27, 
    2268.84, 2220.27, 2185.11, 2136.54, 
    2118.96, 2202.69, 2251.25, 
    2304.0, 2052.82, 2118.96, 2286.42, 
    2136.54, 2118.96, 2220.27, 
    2118.96, 2118.96, 2268.84, 2304.0, 
    2101.38, 2268.84, 2286.42, 
    2035.23, 2304.0, 2202.69, 2118.96, 
    2304.0, 2202.69, 2017.65, 
    2268.84, 2220.27, 2136.54, 2304.0,
    2202.69, 2118.96, 2268.84, 
    2136.54, 2136.54, 2202.69, 2118.96, 
    2035.23, 2202.69, 2304.0, 
    2118.96, 2220.27, 2136.54, 2185.11, 
    2118.96, 2220.27, 2251.25, 
    2220.27, 2304.0, 2202.69, 2202.69, 
    2220.27, 2268.84, 2304.0, 
    2185.11, 2136.54, 2202.69, 2118.96, 
    2185.11, 2220.27, 2118.96, 
    2118.96, 2268.84, 2220.27, 2185.11, 
    2136.54, 2251.25, 2304.0, 
    2268.84, 2202.69, 2220.27, 2185.11, 
    2202.69, 2118.96, 2220.27, 
    2220.27, 2220.27, 2286.42, 2268.84, 
    2220.27, 2185.11, 2202.69, 
    2101.38, 2202.69, 2220.27, 2202.69, 
    2304.0, 2304.0, 2220.27, 
    2202.69, 2220.27, 2286.42, 2220.27, 
    2118.96, 2118.96, 2220.27, 
    2202.69, 2035.23, 2118.96, 2202.69, 
    2286.42, 2118.96, 2101.38, 
    2118.96, 2202.69, 2304.0, 2136.54, 
    2304.0, 2185.11, 2202.69, 
    2286.42, 2101.38, 2202.69, 2220.27, 
    2220.27, 2220.27, 2220.27, 
    2220.27, 2136.54, 2220.27, 2101.38, 
    2220.27, 2286.42, 2118.96, 
    2220.27, 2220.27, 2286.42, 2220.27,
    2136.54, 2136.54, 2268.84, 
    2101.38, 2202.69, 2101.38, 2035.23, 
    2220.27, 2136.54, 2202.69, 
    2136.54, 2286.42, 2185.11, 2286.42, 
    2220.27, 2118.96, 2118.96, 
    2220.27, 2202.69, 2202.69, 2118.96, 
    2118.96, 2118.96, 2202.69, 
    2220.27, 2202.69, 2220.27, 2202.69, 
    2286.42, 2118.96, 2202.69, 
    2136.54, 2220.27, 2268.84, 2118.96,
    2251.25, 2286.42, 2136.54, 
    2136.54, 2286.42, 2286.42, 2220.27,
    2202.69, 2286.42, 2136.54, 
    2220.27, 2202.69, 2220.27, 2118.96,
    2202.69, 2304.0, 2220.27, 
    2202.69, 2304.0, 2286.42, 2304.0,
    2286.42, 2101.38, 2220.27, 
    2101.38, 2220.27, 2202.69, 2202.69,
    2220.27, 2035.23, 2304.0, 
    2304.0, 2202.69, 2118.96, 2185.11,
    2202.69, 2220.27, 2118.96, 
    2101.38, 2202.69, 2268.84, 2304.0,
    2202.69, 2017.65, 2118.96, 
    2286.42, 1933.92, 2304.0, 2118.96, 
    2202.69, 2286.42, 2220.27, 
    2136.54, 2136.54, 2185.11, 2017.65,
    2185.11, 2220.27, 2202.69, 
    2118.96, 2136.54, 2118.96, 2220.27,
    2017.65, 2251.25, 2136.54, 
    2202.69, 2035.23, 1933.92, 2136.54,
    2017.65, 2220.27, 2118.96, 
    2101.38, 2220.27, 2286.42, 2118.96, 
    2286.42, 2220.27, 2286.42, 
    2220.27, 2286.42, 2202.69, 2268.84,
    2101.38, 2251.25, 2220.27, 
    2035.23, 2286.42, 2202.69, 2202.69
\end{verbatim}
The sample mean of the 36 initial samples is: 2193.1, and the sample mean of the extra is: 2192.4. We can see that these are similar values ($0.03\%$ difference), and so we can conclude the method's effectiveness \textit{on average} by only taking 36 samples. This is very useful as minimizing time and resource waste is an important factor when carrying out simulations.

\section[\appendixname~\thesection]{Degeneracy of Solutions} \label{appen:Degen}
Below is a list of all possible optimal solutions for the problem defined in \cref{sec:TM}.
\begin{verbatim}
 (0, 0, 0, 0, 0, 1, 0, 1, 0, 0, 0, 0, 0, 1, 0, 1),
 (0, 0, 0, 0, 0, 1, 0, 1, 0, 0, 0, 0, 1, 0, 0, 1),
 (0, 0, 0, 0, 0, 1, 0, 1, 0, 0, 0, 0, 1, 0, 1, 0),
 (0, 0, 0, 0, 1, 0, 0, 1, 0, 0, 0, 0, 0, 1, 0, 1),
 (0, 0, 0, 0, 1, 0, 0, 1, 0, 0, 0, 0, 1, 0, 0, 1),
 (0, 0, 0, 0, 1, 0, 0, 1, 0, 0, 0, 0, 1, 0, 1, 0),
 (0, 0, 0, 0, 1, 0, 1, 0, 0, 0, 0, 0, 0, 1, 0, 1),
 (0, 0, 0, 0, 1, 0, 1, 0, 0, 0, 0, 0, 1, 0, 0, 1),
 (0, 0, 0, 0, 1, 0, 1, 0, 0, 0, 0, 0, 1, 0, 1, 0),
 (0, 0, 0, 1, 0, 1, 0, 0, 0, 0, 0, 0, 0, 1, 0, 1),
 (0, 0, 0, 1, 0, 1, 0, 0, 0, 0, 0, 0, 1, 0, 0, 1),
 (0, 0, 0, 1, 0, 1, 0, 0, 0, 0, 0, 0, 1, 0, 1, 0),
 (0, 0, 0, 1, 0, 1, 0, 0, 0, 0, 0, 1, 0, 1, 0, 0),
 (0, 0, 0, 1, 0, 1, 0, 0, 0, 0, 0, 1, 1, 0, 0, 0),
 (0, 0, 0, 1, 1, 0, 0, 0, 0, 0, 0, 0, 0, 1, 0, 1),
 (0, 0, 0, 1, 1, 0, 0, 0, 0, 0, 0, 0, 1, 0, 0, 1),
 (0, 0, 0, 1, 1, 0, 0, 0, 0, 0, 0, 0, 1, 0, 1, 0),
 (0, 0, 0, 1, 1, 0, 0, 0, 0, 0, 0, 1, 0, 1, 0, 0),
 (0, 0, 0, 1, 1, 0, 0, 0, 0, 0, 0, 1, 1, 0, 0, 0),
 (0, 0, 0, 1, 1, 0, 0, 0, 0, 0, 1, 0, 1, 0, 0, 0),
 (0, 0, 1, 0, 1, 0, 0, 0, 0, 0, 0, 0, 0, 1, 0, 1),
 (0, 0, 1, 0, 1, 0, 0, 0, 0, 0, 0, 0, 1, 0, 0, 1),
 (0, 0, 1, 0, 1, 0, 0, 0, 0, 0, 0, 0, 1, 0, 1, 0),
 (0, 0, 1, 0, 1, 0, 0, 0, 0, 0, 0, 1, 0, 1, 0, 0),
 (0, 0, 1, 0, 1, 0, 0, 0, 0, 0, 0, 1, 1, 0, 0, 0),
 (0, 0, 1, 0, 1, 0, 0, 0, 0, 0, 1, 0, 1, 0, 0, 0),
 (0, 1, 0, 0, 0, 0, 0, 1, 0, 0, 0, 0, 0, 1, 0, 1),
 (0, 1, 0, 0, 0, 0, 0, 1, 0, 0, 0, 0, 1, 0, 0, 1),
 (0, 1, 0, 0, 0, 0, 0, 1, 0, 0, 0, 0, 1, 0, 1, 0),
 (0, 1, 0, 0, 0, 0, 0, 1, 0, 1, 0, 0, 0, 0, 0, 1),
 (0, 1, 0, 0, 0, 0, 0, 1, 1, 0, 0, 0, 0, 0, 0, 1),
 (0, 1, 0, 0, 0, 0, 0, 1, 1, 0, 0, 0, 0, 0, 1, 0),
 (0, 1, 0, 1, 0, 0, 0, 0, 0, 0, 0, 0, 0, 1, 0, 1),
 (0, 1, 0, 1, 0, 0, 0, 0, 0, 0, 0, 0, 1, 0, 0, 1),
 (0, 1, 0, 1, 0, 0, 0, 0, 0, 0, 0, 0, 1, 0, 1, 0),
 (0, 1, 0, 1, 0, 0, 0, 0, 0, 0, 0, 1, 0, 1, 0, 0),
 (0, 1, 0, 1, 0, 0, 0, 0, 0, 0, 0, 1, 1, 0, 0, 0),
 (0, 1, 0, 1, 0, 0, 0, 0, 0, 0, 1, 0, 1, 0, 0, 0),
 (0, 1, 0, 1, 0, 0, 0, 0, 0, 1, 0, 0, 0, 0, 0, 1),
 (0, 1, 0, 1, 0, 0, 0, 0, 0, 1, 0, 1, 0, 0, 0, 0),
 (0, 1, 0, 1, 0, 0, 0, 0, 1, 0, 0, 0, 0, 0, 0, 1),
 (0, 1, 0, 1, 0, 0, 0, 0, 1, 0, 0, 0, 0, 0, 1, 0),
 (0, 1, 0, 1, 0, 0, 0, 0, 1, 0, 0, 1, 0, 0, 0, 0),
 (0, 1, 0, 1, 0, 0, 0, 0, 1, 0, 1, 0, 0, 0, 0, 0),
 (1, 0, 0, 0, 0, 0, 0, 1, 0, 0, 0, 0, 0, 1, 0, 1),
 (1, 0, 0, 0, 0, 0, 0, 1, 0, 0, 0, 0, 1, 0, 0, 1),
 (1, 0, 0, 0, 0, 0, 0, 1, 0, 0, 0, 0, 1, 0, 1, 0),
 (1, 0, 0, 0, 0, 0, 0, 1, 0, 1, 0, 0, 0, 0, 0, 1),
 (1, 0, 0, 0, 0, 0, 0, 1, 1, 0, 0, 0, 0, 0, 0, 1),
 (1, 0, 0, 0, 0, 0, 0, 1, 1, 0, 0, 0, 0, 0, 1, 0),
 (1, 0, 0, 0, 0, 0, 1, 0, 0, 0, 0, 0, 0, 1, 0, 1),
 (1, 0, 0, 0, 0, 0, 1, 0, 0, 0, 0, 0, 1, 0, 0, 1),
 (1, 0, 0, 0, 0, 0, 1, 0, 0, 0, 0, 0, 1, 0, 1, 0),
 (1, 0, 0, 0, 0, 0, 1, 0, 1, 0, 0, 0, 0, 0, 0, 1),
 (1, 0, 0, 0, 0, 0, 1, 0, 1, 0, 0, 0, 0, 0, 1, 0),
 (1, 0, 0, 1, 0, 0, 0, 0, 0, 0, 0, 0, 0, 1, 0, 1),
 (1, 0, 0, 1, 0, 0, 0, 0, 0, 0, 0, 0, 1, 0, 0, 1),
 (1, 0, 0, 1, 0, 0, 0, 0, 0, 0, 0, 0, 1, 0, 1, 0),
 (1, 0, 0, 1, 0, 0, 0, 0, 0, 0, 0, 1, 0, 1, 0, 0),
 (1, 0, 0, 1, 0, 0, 0, 0, 0, 0, 0, 1, 1, 0, 0, 0),
 (1, 0, 0, 1, 0, 0, 0, 0, 0, 0, 1, 0, 1, 0, 0, 0),
 (1, 0, 0, 1, 0, 0, 0, 0, 0, 1, 0, 0, 0, 0, 0, 1),
 (1, 0, 0, 1, 0, 0, 0, 0, 0, 1, 0, 1, 0, 0, 0, 0),
 (1, 0, 0, 1, 0, 0, 0, 0, 1, 0, 0, 0, 0, 0, 0, 1),
 (1, 0, 0, 1, 0, 0, 0, 0, 1, 0, 0, 0, 0, 0, 1, 0),
 (1, 0, 0, 1, 0, 0, 0, 0, 1, 0, 0, 1, 0, 0, 0, 0),
 (1, 0, 0, 1, 0, 0, 0, 0, 1, 0, 1, 0, 0, 0, 0, 0),
 (1, 0, 1, 0, 0, 0, 0, 0, 0, 0, 0, 0, 0, 1, 0, 1),
 (1, 0, 1, 0, 0, 0, 0, 0, 0, 0, 0, 0, 1, 0, 0, 1),
 (1, 0, 1, 0, 0, 0, 0, 0, 0, 0, 0, 0, 1, 0, 1, 0),
 (1, 0, 1, 0, 0, 0, 0, 0, 0, 0, 0, 1, 0, 1, 0, 0),
 (1, 0, 1, 0, 0, 0, 0, 0, 0, 0, 0, 1, 1, 0, 0, 0),
 (1, 0, 1, 0, 0, 0, 0, 0, 0, 0, 1, 0, 1, 0, 0, 0),
 (1, 0, 1, 0, 0, 0, 0, 0, 0, 1, 0, 0, 0, 0, 0, 1),
 (1, 0, 1, 0, 0, 0, 0, 0, 0, 1, 0, 1, 0, 0, 0, 0),
 (1, 0, 1, 0, 0, 0, 0, 0, 1, 0, 0, 0, 0, 0, 0, 1),
 (1, 0, 1, 0, 0, 0, 0, 0, 1, 0, 0, 0, 0, 0, 1, 0),
 (1, 0, 1, 0, 0, 0, 0, 0, 1, 0, 0, 1, 0, 0, 0, 0),
 (1, 0, 1, 0, 0, 0, 0, 0, 1, 0, 1, 0, 0, 0, 0, 0)
\end{verbatim}
\end{document}